\documentstyle[12pt]{article}

\begin{document}


\newif\iffigureexists
\newif\ifepsfloaded
\openin 1 epsf.sty
\ifeof 1 \epsfloadedfalse \else \epsfloadedtrue \fi
\closein 1
\ifepsfloaded
    \input epsf.sty
\else
    \immediate\write20{>Warning:
         No epsf.sty --- cannot embed Figures!!}
\fi
\def\checkex#1 {\relax
    \ifepsfloaded \openin 1 #1
        \ifeof 1 \figureexistsfalse
        \else \figureexiststrue
        \fi \closein 1
    \else \figureexistsfalse
    \fi }

\def\epsfhako#1#2#3#4#5#6{
\checkex{#1}
\iffigureexists
    \begin{figure}[#2]
    \epsfxsize=#3
    \centerline{\epsffile{#1}}
    {#6}
    \caption{#4}
    \label{#5}
    \end{figure}
\else
    \begin{figure}[#2]
    \caption{#4}
    \label{#5}
    \end{figure}
    \immediate\write20{>Warning:
         Cannot embed a Figure (#1)!!}
\fi
}

\ifepsfloaded
\checkex{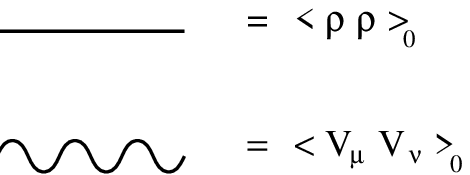}
    \iffigureexists \else
    \immediate\write20{>EPS files for Figs. 2 and 3 are packed
     in a uuecoded compressed tar file}
    \immediate\write20{>appended to this LaTeX file.}
    \immediate\write20{>You should unpack them and LaTeX again!!}
    \fi
\fi

\renewcommand{\thefootnote}{\fnsymbol{footnote}}
\renewcommand{\theequation}{\thesection.\arabic{equation}}
\newcommand{\reseteqnum}{\setcounter{equation}{0}}

\newcommand{\hako}{\lower1pt\hbox{$\Box$}}
\newcommand{\D}{\partial}
\newcommand{\fr}{\frac}
\newcommand{\FF}{F_{\mu\nu}F_{\mu\nu}}
\newcommand{\VV}{V_{\mu\nu}V_{\mu\nu}}
\newcommand{\ff}{f_{\mu\nu}f_{\mu\nu}}
\newcommand{\JJ}{J_{\mu\nu}(x)J_{\mu\nu}(x)}
\newcommand{\Vf}{V_{\mu\nu}f_{\mu\nu}}
\newcommand{\Bf}{\epsilon_{\mu\nu\rho\sigma}B_{\mu\nu}f_{\rho\sigma}}
\newcommand{\BJ}{B_{\mu\nu}J_{\mu\nu}}
\newcommand{\FtF}{F_{\mu\nu}\tilde{F}_{\mu\nu}}
\newcommand{\VtV}{V_{\mu\nu}\tilde{V}_{\mu\nu}}
\newcommand{\Vtf}{V_{\mu\nu}\tilde{f}_{\mu\nu}}
\newcommand{\ftf}{f_{\mu\nu}\tilde{f}_{\mu\nu}}
\newcommand{\JtJ}{J_{\mu\nu}\tilde{J}_{\mu\nu}}
\newcommand{\dtds}{d^{2}\sigma}
\newcommand{\dtdsp}{d^{2}\sigma'}
\newcommand{\dx}{d^{4}x}
\newcommand{\ep}{\epsilon_{\mu\nu\rho\sigma}}
\newcommand{\eps}{\epsilon_{\mu\nu\rho\sigma}}
\newcommand{\epsi}{\epsilon_{ijk}}
\newcommand{\TH}{\Theta(\mbox{\boldmath $x$}-\mbox{\boldmath $X$})}
\newcommand{\tiF}{\frac{1}{2}\epsilon_{\mu\nu\rho\sigma}F_{\rho\sigma}}
\newcommand{\tF}{\tilde{F}}
\newcommand{\tV}{\tilde{V}}
\newcommand{\tf}{\tilde{f}}
\newcommand{\tJ}{\tilde{J}}
\newcommand{\s}{\sigma}
\newcommand{\Si}{\Sigma}
\newcommand{\th}{\theta}
\newcommand{\Th}{\Theta}
\newcommand{\r}{\rho}
\newcommand{\la}{\lambda}
\newcommand{\La}{\Lambda}
\newcommand{\m}{\mu}
\newcommand{\n}{\nu}
\newcommand{\ta}{\tau}
\newcommand{\de}{\delta}
\newcommand{\De}{\Delta}
\newcommand{\ga}{\gamma}
\newcommand{\Ga}{\Gamma}
\newcommand{\e}{\epsilon}
\newcommand{\ve}{\varepsilon}
\newcommand{\XX}{\fr{\D X_{[\mu}}{\D \tau}\fr{\D X_{\nu]}}{\D \sigma}}
\newcommand{\YY}{\fr{\D Y_{[\mu}}{\D \tau}\fr{\D Y_{\nu]}}{\D \sigma}}
\newcommand{\rXX}{\fr{\D X_{[\rho}}{\D \tau'}\fr{\D X_{\sigma]}}
{\D \sigma'}}
\newcommand{\rYY}{\fr{\D Y_{[\rho}}{\D \tau'}\fr{\D Y_{\sigma]}}
{\D \sigma'}}
\newcommand{\Dd}{\de^{(4)}}
\newcommand{\bX}{\mbox{\boldmath $X$}}
\newcommand{\bx}{\mbox{\boldmath $x$}}
\newcommand{\by}{\mbox{\boldmath $y$}}
\newcommand{\bY}{\mbox{\boldmath $Y$}}
\newcommand{\bn}{\mbox{\boldmath $n$}}
\newcommand{\bR}{\mbox{\boldmath $R$}}
\newcommand{\be}{\mbox{\boldmath $e$}}
\newcommand{\bp}{\mbox{\boldmath $p$}}
\newcommand{\bP}{\mbox{\boldmath $P$}}
\newcommand{\bZ}{\mbox{\boldmath $Z$}}
\newcommand{\bE}{\mbox{\boldmath $E$}}
\newcommand{\bB}{\mbox{\boldmath $B$}}
\newcommand{\sD}{\partial\hspace{-0.5em}/}
\newcommand{\sA}{A\hspace{-0.55em}/}

\renewcommand{\thepage}{ }

\begin{titlepage}
\title{
\hfill
\parbox{4cm}{\normalsize KUNS-1269\\HE(TH)~94/08\\hep-th/9406208}\\
\vspace{1cm}
``Topological'' Formulation of Effective Vortex Strings}
\author{Masatoshi Sato\thanks{e-mail address:
\tt{msato@gauge.scphys.kyoto-u.ac.jp}}{\,\,}\thanks{Fellow of the Japan
Society for the Promotion of Science for Japanese Junior Scientists.}
 and Shigeaki Yahikozawa\thanks{e-mail address:
\tt{yahiko@gauge.scphys.kyoto-u.ac.jp}}\\
{\normalsize\em Department of Physics, Kyoto University}\\
{\normalsize\em Kyoto 606-01, Japan}}
\date{\normalsize June, 1994}
\maketitle

\begin{abstract}
\normalsize
We present a ``topological'' formulation of arbitrarily shaped vortex
strings in four dimensional field theory.
By using a large Higgs mass expansion, we then evaluate the effective
action of the closed Abrikosov-Nielsen-Olesen vortex string.
It is shown that the effective action contains the Nambu-Goto term
and an extrinsic curvature squared term with negative sign.
We next evaluate the topological $\FtF$ term and find that it becomes the
sum of an ordinary self-intersection number and Polyakov's
self-intersection number of the world sheet swept by the vortex string.
These self-intersection numbers are related to the self-linking number
and the total twist number, respectively.
Furthermore, the $\FtF$ term turns out to be the difference
between the sum of the writhing numbers and the linking numbers of the
vortex strings at the initial time and the one at the final time.
When the vortex string is coupled to fermions, the chiral fermion
number of the vortex string becomes the writhing number (modulo
$\bZ$) through the chiral anomaly.
Our formulation is also applied to ``global'' vortex strings in a
model with a broken global $U(1)$ symmetry.
\end{abstract}
\end{titlepage}

\newpage
\renewcommand{\thepage}{\arabic{page}}
\setcounter{page}{1}

\baselineskip=20.5pt plus 0.2pt minus 0.1pt

\section{Introduction}\label{sec:Int}

The study of string-like objects has been actively pursued from both
theoretical and experimental interests in various
fields including condensed matter physics and biology.
In particle physics and cosmology, the topological vortex string
arising in field theory is one of the most interesting string-like objects.
In particular, the Abrikosov-Nielsen-Olesen (ANO) vortex string is the
simplest one that splendidly shows typical properties of the vortex
string \cite{NO}.
Toward a better understanding of the physics on the vortex string, it
is important to examine its geometric and topological properties in
four space-time dimensions.
We are especially interested in four dimensional extrinsic properties
such as the entanglement of the vortex strings.
For studying them, we need a systematic formulation of the vortex
string.
The method used so far in evaluating the effective action of the
vortex string in arbitrary shape is based on F\"orster's
parameterization of coordinates \cite{Fo} and the collective
coordinates method \cite{GS}.
In this method, however, the cut-off dependence of the theory is not
so clear and topological structures such as the self-intersection of
the world sheet swept by the vortex string cannot be so easily
investigated.
Therefore, it is desirable to construct a more systematic and efficient
formulation satisfying the following: (i) changes of the shapes of
the vortex strings can be described, (ii) using perturbative expansions
by appropriate parameters such as coupling constants, masses or
cut-offs, one can perform systematic approximations, (iii) topological
features of the vortex string can be easily examined.

   In this paper, we present a ``topological'' formulation which
satisfies the above three requirements and apply it to the arbitrarily
shaped vortex strings in field theories with broken local or global
$U(1)$ symmetries.
This is a relativistic generalization of the ``topological'' formulation
used in the study of quantized vortices in superfluid helium \cite{HYAT}.
One of the characteristic features in our formulation is the appearance
of an antisymmetric tensor field and a so-called topological BF term
\cite{BBRT}.
We also adopt a manifestly Lorentz invariant Gaussian-type
regularization for the $\de$-functions in vorticity tensor currents.
Using our formulation, we first evaluate the effective action of
the ANO vortex string in the Abelian Higgs
model, showing that it contains not only the Nambu-Goto term but also
an extrinsic curvature squared term with negative sign.
Second, we examine the topological $\FtF$ term,
which for example appears as the chiral anomaly and the $\theta$ term.
The evaluation of this term tells us that there are interesting
relations between several geometric or topological quantities:
Polyakov's self-intersection number, ordinary self-intersection
number, total twist number, self-linking number,
writhing number and linking number.
The expectation value of the $\FtF$ term turns out
to be the sum of Polyakov's self-intersection number and the ordinary
self-intersection number of the world sheet swept by the vortex
string at the leading order of our approximation.
In addition, the $\FtF$ term can be written as the difference of the
sum of the writhing number and the linking number at the final time
and the one at the initial time.
Furthermore, we discuss the chiral fermion number of the ANO vortex
strings in arbitrary shape by using the chiral anomaly and find
it to be the sum of the writhing numbers of each vortex string (modulo
$\bZ$).
To make the validity of our formulation clearer, we also study the
dynamics of vortex strings in a model with a broken global $U(1)$
symmetry. (In cosmology, the former ANO vortex strings are called
``local strings'' and the latter ones are called ``global strings''.)
In both models, it is also shown that the large Higgs mass expansions are
good approximations.
As a whole, it is demonstrated that our ``topological''
formulation is useful to study the ``effective'' vortex strings.
(The ``effective'' vortex string means the vortex string remaining
after integration over massive fields in field theory.)

Our formulation can be applied to several phenomenologies, although
we do not completely discuss them in this paper.
First, when we consider grand unified models with extra broken $U(1)$
symmetries, then there can exist vortex strings
which are topologically stable.
Their string tension is
of order a GUT-scale squared or
possibly a fundamental string scale squared (because the gauge
coupling is smaller than 1).
Furthermore, when the models have anomalous $U(1)$ symmetries,
it is interesting to investigate whether any fermion number
can be violated through the effect of the vortex string.
Second, in the Weinberg-Salam theory, there appears the so-called
$Z$ string which is equivalent to the ANO vortex string if one neglects
other degrees of freedom \cite{Na,Va}.
It is a kind of sphaleron \cite{KO}, which perhaps seems to be related
to the weak-scale baryogenesis through the chiral anomaly \cite{tH,KRS}.
Therefore, the investigation of the vortex string in arbitrary shape is
important from the point of view of the fermion number violation.
Third, our theory can be directly applied to the cosmic string model
\cite{Vi} and superconductor systems.
Finally, it should be noticed that the study of the effective vortex
string would give us a new angle in understanding extrinsic properties
of fundamental strings in four dimensional space-time.

The paper is organized as follows:
in sect. 2 and 3, we study the Abelian Higgs model with the vortex string.
First, in sect. 2, we present our ``topological'' formulation and evaluate
the effective action of the vortex string.
Next, in sect. 3, the $\FtF$ term is examined
and geometric or topological relations are shown. We also discuss
the chiral fermion number of the vortex string in arbitrary shape.
In sect. 4, we examine a model with the broken global $U(1)$
symmetry.
In sect. 5, we give conclusions and compare our results with previous ones.
In appendix A, we explain the relation between Polyakov's
self-intersection number and the total twist number.
In appendix B, we derive the relation between the
intersection number and the linking number.

\section{``Topological'' formulation and the effective action of the
ANO vortex string}\label{sec:Loc}
\reseteqnum

In this section, we consider the Abelian Higgs model with an
arbitrarily shaped vortex string in four space-time dimensions.
Furthermore, for simplicity we suppose the vortex string to be a closed
one with circulation number one. It is easy to extend to the case with
many vortex strings with arbitrary circulation numbers.
We always use the Euclidean formulation of field theory,
so the model is described by the Lagrangian
\begin{eqnarray}
{\cal L}=\frac{1}{4}\FF
   +|(\D_{\mu}-ieA_{\mu})\varphi|^{2}
    +2\lambda\biggl(\varphi^{\dagger}\varphi-\frac{\eta^{2}}{2}\biggr)^{2},
\label{eqn:AH}
\end{eqnarray}
where $A_{\mu}$ is an ordinary $U(1)$ gauge field, $\varphi$
a complex scalar field and $F_{\mu\nu}$ a field strength
tensor of $A_{\mu}$.
The existence of the vortex string with circulation one means that
when one takes a turn along any closed contour around the vortex
string, the phase of the scalar field changes by 2$\pi$. To get such
a phase, we use the solid angle subtended by the vortex string
\begin{equation}
\theta(\mbox{\boldmath $x$};\mbox{\boldmath $X$})
 =\frac{1}{2}\int_{S} d\mbox{\boldmath $S$}'\cdot\nabla
 \frac{1}{|\mbox{\boldmath $x$}-\mbox{\boldmath $x$}'|},
\label{eqn:Sol}
\end{equation}
where $S$ is any surface bounded by the vortex string: $\partial{S}=
\Gamma$, where $\Gamma=\{\mbox{\boldmath$X$}(\sigma_1,t); 0\leq\sigma_1
\leq2\pi\}$ and $\mbox{\boldmath$X$}(\sigma_1,t)$ denotes the position
of the vortex string at time $t$.
In terms of this solid angle, the scalar field with the vortex string
is described by
\begin{equation}
\varphi(x)=\exp\{i\theta(\mbox{\boldmath $x$};
\mbox{\boldmath$X$})\}\phi(x),
\label{eqn:Tra}
\end{equation}
where $\phi(x)$ is a regular function.
When we substitute $\varphi(x)$ in (\ref{eqn:AH}), the differential
term of $\varphi(x)$ changes and the covariance of (\ref{eqn:AH})
is apparently broken.
So we need an alternative formulation where the covariance is manifest
and it is easier to deal with the vortex string.

We propose a manifestly covariant Lagrangian with a topological term
and a vorticity tensor current,
which is equivalent to the original one (\ref{eqn:AH}) in the sense
explained later:
\begin{equation}
{\cal L}=\frac{1}{4}\FF
   +|(\D_{\mu}-ieA_{\mu}-ia_{\mu})\phi|^{2}
   +2\lambda\biggl(\phi^{\dagger}\phi-\frac{\eta^{2}}{2}\biggr)^{2}
   +i\Bf +iB_{\mu\nu}J_{\mu\nu},
\label{eqn:TAH}
\end{equation}
where
\begin{equation}
J_{\mu\nu}(x)=-4\pi\int\dtds
\partial_{1}X_{[ \mu} \partial_{2}X_{\nu ]}
\delta^{(4)}(x-X(\sigma))
\label{eqn:Cur}
\end{equation}
is the vorticity tensor current and
$a_{\mu}$ another $U(1)$ gauge field for the vorticity, $B_{\mu\nu}$ a
rank two antisymmetric tensor field, $f_{\mu\nu}$ a field strength
tensor of $a_{\mu}$,
$\partial_{a}=\partial / \partial\sigma_{a}$ $(a=1,2)$
and $A_{[\mu}B_{\nu]}=A_{\mu}B_{\nu}-A_{\nu}B_{\mu}$.
Here $X_{\mu}(\sigma)$ denotes the four dimensional location
of the vortex string,
where $\sigma=(\sigma_1, \sigma_2)$ are the coordinates which
parameterize the world sheet swept by the vortex string.

 Actually, variations of (\ref{eqn:TAH}) with respect to $B_{\mu\nu}$
lead to the constraint
$\epsilon_{\mu\nu\rho\sigma}f_{\rho\sigma}+J_{\mu\nu}=0$ and if we
choose the Coulomb gauge $\partial_{i}a_{i}=0$ and
a gauge $\sigma_{2}=X_{4}=t$, then we get $a_{\mu}=-\partial_{\mu}
\theta$ and
\begin{equation}
\left(\partial_\mu-ieA_\mu\right)\varphi=
\left(\partial_\mu-ieA_\mu-ia_\mu\right)\phi ,
\label{eqn:Der}
\end{equation}
so that this Lagrangian (\ref{eqn:TAH}) turns out to be the original one
(\ref{eqn:AH}) into which the redefined scalar field $\varphi(x)$ is
inserted.
{}From now on, let us use the Lagrangian (\ref{eqn:TAH})
as our starting point for the theory.

The Lagrangian (\ref{eqn:TAH}) has some interesting properties.
(i) It has two types of gauge symmetries except the usual $U(1)$
symmetry. The first one is another $U(1)$ gauge symmetry:
$a_{\mu}\rightarrow a_{\mu}+\partial_{\mu}\alpha$ and
$\phi\rightarrow e^{i\alpha}\phi$ with an arbitrary regular function
$\alpha$. The second one is given by $B_{\mu\nu}\rightarrow B_{\mu\nu}
+\partial_{\mu}\Lambda_{\nu}-\partial_{\nu}\Lambda_{\mu}$, where
$\Lambda_{\mu}$ is also an arbitrary regular function.
The corresponding conserved tensor current is $J_{\mu\nu}$, so that the total
vorticity $\int d^{3}x J_{4i}$ is conserved.
(ii) It has the topological term $\Bf$,
which is called a BF term.
In general the BF term is used in evaluating linking numbers
which are topological numbers counting how many times a string
and a two dimensional membrane are entangled in four dimensions \cite{BBRT}.
The topological BF term is a generalization of the Chern-Simons term
which plays an important role in the study of the quantized Hall
effect and anyon systems in $2+1$ dimensions \cite{WZ}.
Furthermore, this term appears in
various areas of theoretical physics, for example, in models with
anomalous U(1) charges in superstring theory \cite{LNS} and four
dimensional 2-form gravity \cite{CDJM}.
Through this term, we may be
able to search some connections between such theories and the present
theory on the vortex string.
(iii) Since the vortex string coordinates
$X_{\mu}(\sigma)$ are contained only in the vorticity tensor current
$J_{\mu\nu}$, which can describe the vortex string in arbitrary shape,
our formulation is useful in many cases, for example, in cases where
the vortex core needs regularizing or one derives the equation
of motion of the vortex string.

We are interested in the effective action of the vortex string defined
by
\begin{equation}
{\cal S}_{eff}[X]= - \ln\int{\cal D}\phi{\cal D}A_{\mu}{\cal D}
                     a_{\mu}{\cal D}B_{\mu\nu}
                     \exp\left(-\int d^{4}x {\cal L}\right).
\label{eqn:Eff}
\end{equation}
This path integral representation is suitable for the evaluation of
the effective action since systematic perturbative expansions can be
easily performed.
We do not consider the quantization of $X_{\m}(\s)$ and the
effects from loops of the fields in this
paper, although it is of course interesting to study them.
Before computing the effective action, let us discuss some points to
clarify our procedure.

As well known, in the static solution of the straight ANO vortex, it
is satisfied that $\phi - \eta/\sqrt{2} \approx 0$ and
$eA_{i}+a_{i}\approx 0$ in the distant region from the vortex core, where
$a_{i}=-\D_{i}\theta_{s}$ ($\theta_{s}$ is an azimuthal angle)
and ``$\approx 0$" implies ``exponentially small".
Even in general cases with moving vortex strings, if we require that
the energy is finite, it should be satisfied that
$\phi - \eta/\sqrt{2} \approx 0$ and
$eA_{\mu}+a_{\mu}\approx 0$ in the distant region from the vortex core.
Furthermore, since $\pi_{1}(U(1))=\mbox{\boldmath $Z$}$,
the vortex string is topologically stable. It is therefore
reasonable to expand $\phi$ and $eA_{\mu}+a_{\mu}$ around
$\eta/\sqrt{2}$ and $0$ respectively in such a region.
So, instead of $\phi$ and $A_{\mu}$,
let us adopt two scalar fields $\rho$, $\omega$ and a vector field
$C_{\mu}$ defined by
\begin{eqnarray}
&&\phi(x)=\frac{1}{\sqrt{2}}(\eta+\rho(x))e^{i\omega(x)},\label{eqn:rho}\\
&&C_{\mu}(x)=eA_{\mu}(x)+a_{\mu}(x).
\label{eqn:C}
\end{eqnarray}
After these replacements, our action still keeps the gauge
invariance under $a_{\mu}\rightarrow
a_{\mu}+\D_{\mu}\alpha$, so that we get a gauge invariant effective
action for $a_{\mu}$ when we integrate over
the fields $\rho$, $\omega$ and $C_{\mu}$.
This fact is convenient for evaluating the effective action of the
vortex string.

In the unitary gauge, where $C_{\mu}-\D_{\mu}\omega$ is replaced
with $V_{\mu}$, the Lagrangian becomes
\begin{eqnarray}
{\cal L}&=&\frac{1}{4e^{2}}\VV+\frac{1}{2}\eta^{2}V_{\mu}
        V_{\mu}\nonumber\\
        & &+\frac{1}{2}\D_{\mu}\rho\D_{\mu}\rho+2\lambda
	\eta^{2}\rho^{2} \nonumber\\
        & &-\frac{1}{2e^{2}}\Vf+\eta\rho V_{\mu}V_{\mu}
           +\frac{1}{2}\rho^{2}V_{\mu}V_{\mu}+2\lambda\eta\rho^{3}
           +\frac{1}{2}\lambda\rho^{4}\nonumber\\
        & &+\frac{1}{4e^{2}}\ff+i\Bf+i\BJ,
\label{eqn:UAH}
\end{eqnarray}
where $V_{\mu\nu}=\D_{\mu}V_{\nu}-\D_{\nu}V_{\mu}$. Propagators of
$\rho$ and $V_{\mu}$, which are denoted as $\langle\rho(x)
\rho(y)\rangle_{0}$ and
$\langle V_{\mu}(x) V_{\nu}(y)\rangle_{0}$ respectively, are given by
\begin{equation}
\langle\rho(x)
\rho(y)\rangle_{0}=\int\frac{d^{4}k}{(2\pi)^{4}}\frac{1}{k^{2}+m_{H}^{2}}
                   e^{ik\cdot(x-y)},
\label{eqn:Pr-r}
\end{equation}
\begin{equation}
\langle V_{\mu}(x) V_{\nu}(y)\rangle_{0}=e^{2}\int\frac{d^{4}k}{(2\pi)^{4}}
        \left(g_{\mu\nu}+\frac{k_{\mu}k_{\nu}}{m_{V}^{2}}\right)
                \frac{1}{k^{2}+m_{V}^{2}}e^{ik\cdot(x-y)},
\label{eqn:Pr-V}
\end{equation}
where $m_{H}^{2}=4\lambda\eta^{2}$ and $m_{V}^{2}=e^{2}\eta^{2}$.
The resulting Feynman rules for the propagators and the vertices are
summarized in fig. \ref{fig:PG} and fig. \ref{fig:VT}.
\epsfhako{propagator.eps}{t}{5cm}{Graphical representation of the
propagators of $\r(x)$ and $V_{\m}(x)$.}{fig:PG}{}
\epsfhako{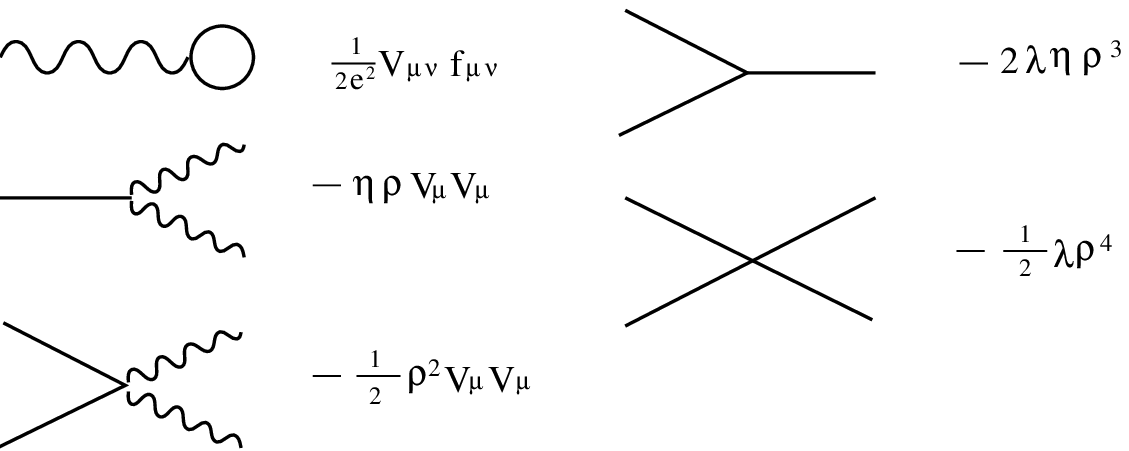}{t}{12cm}{Graphical representation of the
vertices in the Lagrangian (2.10). The circle indicating $f_{\m\n}$
can be actually replaced with the vorticity tensor current
$-2\tJ_{\m\n}$ through the constraint $f_{\m\n}=-2\tJ_{\m\n}$, where
$\tJ_{\m\n}=\fr{1}{2}\eps J_{\r\s}$.}{fig:VT}{}

Note that since we do not exactly know the inside structure of the
vortex core when the vortex string is moving, it is necessary
to assume its structure and introduce an appropriate cut-off
parameter.
The ansatz which we actually apply to our system is that the
$\delta$-function in the vorticity tensor current $J_{\mu\nu}$
is regularized smoothly in the cut-off region.
In this paper, we make use of the first expression of
the following Gaussian-type regularization:
\begin{eqnarray}
\delta_{\Lambda}^{(4)}(x)&=&\frac{\Lambda^{4}}{\pi^{2}}
                            \exp(-\Lambda^{2}x^{2})\nonumber\\
                &=&\int\frac{d^{4}k}{(2\pi)^{4}}
                \exp\left(ik\cdot x-\frac{k^{2}}{4\Lambda^{2}}\right),
\label{eqn:Gauss}
\end{eqnarray}
where the second expression shows that the momentum is effectively
cut off at about $\Lambda$.
In this regularization, the Lorentz invariance is manifest and
the conservation of the vorticity tensor current, $\D_{\mu}J_{\mu\nu}
=0$, is also preserved. It is one of crucial reasons why our formulation is
a convenient one to evaluate
fine structures of the effective action of the vortex string and
topological properties such as the self-intersection number of
the world sheet swept by the vortex string.

In systematic estimations of the order of each tree diagram for the
effective action, we can perform the large mass expansion by powers of
$1/m_{H}^{}$ when $m_{H}^{}$ is larger than other mass scales.
Especially, in this large Higgs mass expansion, the propagator of the Higgs
field $\rho$ can be treated as
\begin{equation}
\langle\rho(x) \rho(y)\rangle_{0} \approx \frac{1}{m_{H}^2} \delta^{(4)}(x-y).
\label{eqn:Pr-r-L}
\end{equation}
Using ordinary relations between numbers of vertices, propagators and
external lines for the diagrams, a simple power counting tells us that
at the tree level the leading power of $1/m_{H}^{}$ for Feynman
diagrams with $N$ circle ($N\ge 2$), which are corresponding to parts
of the effective action with $N$ vorticity tensor currents, is given by
\begin{equation}
\left(\frac{M}{m_{H}^{}}\right)^{N-2}.
\label{eqn:PC}
\end{equation}
Here M are $m_{V}$, $\Lambda$ or $1/R$, where $R$ denotes the
characteristic length which represents the smoothness of the vortex
string, that is, the magnitude of higher order derivatives of $X_{\mu}$.
Hence, it turns out that the diagrams with smaller numbers of
vorticity tensor currents
are dominant for the effective action.

Let us turn our attention now to the evaluation of the effective
action of the vortex string.
The leading contribution ${\cal S}_{0}$ to it, which has two
vorticity tensor currents and corresponds to Feynman diagrams depicted in
fig. \ref{fig:LD}, can be evaluated in the form
\begin{equation}
{\cal S}_{0}=\frac{m_{V}^{2}}{16e^{2}}\int d^{4}x d^{4}y
            J_{\mu\nu}(x)D(x-y)J_{\mu\nu}(y),
\label{eqn:L-Lead}
\end{equation}
where
\begin{equation}
D(x-y)=\int\frac{d^{4}k}{(2\pi)^{4}}
           \frac{1}{k^{2}+m_{V}^{2}}e^{ik\cdot(x-y)}.
\label{eqn:Pr-D}
\end{equation}
Here we have used the conservation law of the vorticity tensor
current $J_{\mu\nu}$.
The form of ${\cal S}_{0}$ expresses that massive particles propagate
between two points on the world sheet swept by the vortex string.
\epsfhako{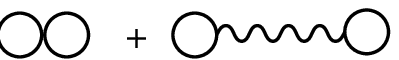}{htb}{4cm}{The diagrams of the leading
contribution to the effective action. Touching circles come from the
$1/4e^{2}\cdot f_{\mu\nu}f_{\mu\nu}$ term in (2.10).}{fig:LD}{}

We will now evaluate the dominant effects arising in the case where
two vorticity tensor currents approach each other. For that purpose,
we adopt the proper time representation of the propagator $D(x-y)$:
\begin{equation}
\frac{1}{k^{2}+m^{2}}=\int_{0}^{\infty}ds \exp[-s(k^{2}+m^{2})].
\label{eqn:ProT}
\end{equation}
After inserting it into ${\cal S}_{0}$ and integrating over $x$, $y$
and $k$, then we find that
\begin{eqnarray}
{\cal S}_{0}&=&\frac{m_{V}^{2}}{16e^{2}}\int_{1/{2 \Lambda^{2}}}^{\infty}
               ds\frac{1}{s^{2}}\int d^{2}\sigma d^{2}{\sigma}'
               \D_{1}X_{[\mu}\D_{2}X_{\nu]}(\sigma)\cdot
               \D_{1}X_{[\mu}\D_{2}X_{\nu]}(\sigma ') \nonumber\\
            & &\qquad \times \exp\left\{ -\frac{1}{4s}
               |X(\sigma)-X(\sigma ')|^{2}
                -sm_{V}^{2}+\frac{m_{V}^{2}}{2\Lambda^{2}}\right\}.
\label{eqn:L-Lead-2}
\end{eqnarray}
Here note that we have used the Gaussian-type regularization of the
$\delta$-function (\ref{eqn:Gauss}).
The factor $e^{-sm^{2}_{V}}$ in (\ref{eqn:L-Lead-2}) indicates that
the region where $s$ is small mainly contributes to ${\cal S}_{0}$ and
the factor $e^{-1/4s\cdot|X(\s)-X(\s')|^{2}}$ in (\ref{eqn:L-Lead-2})
shows that the region where $X(\s)\approx X(\s')$ is the dominant part
of contributions to ${\cal S}_{0}$ when $s$ is small.
Putting these together, it turns out that the dominant contribution
to ${\cal S}_{0}$ comes from the region where $X(\s)$ is near
$X(\s')$, so let us consider the case where $\s'$ is near $\s$.
In order to investigate the behavior of ${\cal S}_{0}$ in the case
where $\sigma '$ is near $\sigma$, we define $z$ as
$z=\sigma '-\sigma$ and expand $X_{\mu}(\sigma ')$ in powers of $z$:
\begin{equation}
X_{\mu}(\sigma ')=X_{\mu}(\sigma)+z_{a}\D_{a}X_{\mu}(\sigma)
                  +\frac{1}{2}(z_{a}\D_{a})^{2}X_{\mu}+ \cdots .
\label{eqn:exp-X}
\end{equation}
After substitution of (\ref{eqn:exp-X}) in (\ref{eqn:L-Lead-2}) and
integration over $z$,
the effective action ${\cal S}_{0}$ takes the particularly simple form
up to $O(1)$ in powers of $(m_{V}^{} R)^{-1}$ and $(\Lambda R)^{-1}$:
\begin{equation}
{\cal S}_{0}=\mu_{0}\int d^{2}\sigma\sqrt{g}
           +\alpha_{0}\int d^{2}\sigma\sqrt{g}K^{A}_{ab}K^{A}_{ab},
\label{eqn:NG}
\end{equation}
where $g_{ab}=\D_{a}X_{\mu}\D_{b}X_{\mu}$ and $g=\det(g_{ab})$.
Here
\begin{equation}
\mu_{0}=\frac{\pi m_{V}^{2}}{2e^{2}}\int_{0}^{\infty}du
    \frac{e^{-u}}{u+\frac{m_{V}^{2}}{2\Lambda^2}},
\label{eqn:ST}
\end{equation}
which is the tension in the vortex string, and
\begin{equation}
\alpha_{0}=-\frac{3\pi}{8e^{2}}.
\label{eqn:K2}
\end{equation}
$K_{ab}^{A}$ is the extrinsic curvature defined by the equation
 \begin{equation}
\D_{a}\D_{b}X_{\mu}=\Gamma^{c}_{ab}\D_{c}X_{\mu}+K^{A}_{ab}n_{\mu}^{A},
\label{eqn:EC}
\end{equation}
where $n_{\mu}^{A}$ are two normal unit vectors perpendicular to
$\D_{a}X_{\mu}$, satisfying $n_{\mu}^{A}n_{\nu}^{B}=\delta^{AB}$ and
$n_{\mu}^{A}\D_{a}X_{\mu}=0$ ($A=1,2$).
Here we have neglected contributions from boundaries and intersection
points of the world sheet swept by the vortex string.
In evaluating the second term of (\ref{eqn:NG}), we have used
the fact that the scalar curvature defined by $g_{ab}$ is a total
divergence.
The first term in the right-hand side of (\ref{eqn:NG}) is the
Nambu-Goto action and the second term is the extrinsic curvature
squared term which was
investigated from the point of view of the QCD string \cite{Po1}.

Let us make a comparison between the string tension $\mu_{0}$ and the one
$\mu$ evaluated by using the static ANO vortex solution in the case where
$m_{H}^{}$ is much larger than $m_{V}^{}$.
Outside the vortex core,
one can treat $\phi$ as a constant $\eta/\sqrt{2}$ and
the ANO vortex solution with circulation one is given by
\begin{eqnarray}
&&A_{i}(x)=\epsilon_{ij}\frac{x_{j}}{r^{2}}\left\{\frac{1}{e}
      -\eta r K_{1}(e\eta r)\right\}\quad(i=1,2),\\
&&A_{3}(x)=0,
\label{eqn:ANO}
\end{eqnarray}
where $K_{1}$ is the modified Bessel function and
$r=\sqrt{x_{1}^{2}+x_{2}^{2}}$ denotes the distance from the center of
the vortex string \cite{NO,Ab}.
Here the ANO vortex solution $A_{i}(x)$ is regular for $r\to0$.
Using a cut-off $\Lambda'$ such that $m_H^{}\gg \Lambda' \gg m_V^{}$,
one can easily obtain the string
tension $\mu$, which is the energy per unit length along the third
axis:
\begin{equation}
\mu \approx \pi\eta^2\ln\left(\frac{\Lambda'}{m_V^{}}\right).
\label{eqn:ST-2}
\end{equation}
The dominant contribution to the string tension $\m$ comes from the
intermediate region satisfying $1/\La'<r< 1/m_{V}$.
In addition, the contribution from the vortex core, which is about $\eta^2$,
is much smaller than the dominant one.
On the other hand, our string tension
$\mu_0$ behaves like
\begin{equation}
\mu_0 \approx \pi\eta^2\ln\left(\frac{\Lambda}{m_V^{}}\right),
\label{eqn:ST-3}
\end{equation}
when $\Lambda\gg m_V^{}$. We therefore get $\Lambda\approx\Lambda'$
if we require both string tensions are equal.
This shows that our formulation is justified when we take about
$\Lambda'$ as $\Lambda$.
Note that we should not use the expansion in positive powers of
$\Lambda/m_{V}^{}$ because $\La\gg m_{V}$, though it is valid to
expand the theory in powers
of $1/m_{H}^{}$ when $m_{H}^{}$ is large enough. This fact is one of
our grounds for using the expansion in powers of $1/m_{H}^{}$.
In addition, it may be worth pointing out that the exponential
integral in the string tension $\m_{0}$ given by (\ref{eqn:ST}) depends on
only the ratio of $m_{V}^{2}$ to $2\La^{2}$.

As concerns the coefficient $\alpha_{0}$, there are two remarkable
points. First, it takes the non-zero value with negative sign,
which indicates that the
vortex string prefers curving as far as it is smooth enough.
Our approximation is relevant only in the case where the
characteristic length representing the curvature of the vortex string
is large enough in comparison with the lengths $1/m_{H}^{}$,
$1/m_{V}^{}$ and $1/\Lambda$, so that the apparent unboundedness
of the effective action (\ref{eqn:NG}) is not a trouble within
the approximation which we have used.
One may notice that our theory is positive definite by
the definition of the Lagrangian (\ref{eqn:TAH}).
Second, the $\alpha_{0}$ is independent of $m_{V}^{}$ and $\Lambda$
unlike the string tension $\mu_{0}$.

To confirm the validity of the expansion by the large Higgs mass,
let us evaluate the next leading contribution whose diagram has four
vorticity tensor currents, as illustrated in fig. \ref{fig:NLD}.
\epsfhako{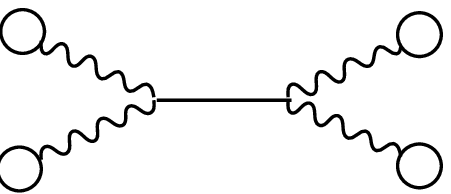}{t}{4.5cm}{The diagram of the next leading
contribution to the effective action.}{fig:NLD}{}
The next leading effective action ${\cal S}_{1}$ is dominated by the
case where the four vorticity tensor currents approach in the same
time, and the dominant term becomes the Nambu-Goto term.
Indeed, by repeating the procedure
used previously, we can calculate ${\cal S}_{1}$:
\begin{equation}
{\cal S}_{1}=\mu_{1}\int d^{2}\sigma\sqrt{g}+\cdots,
\label{eqn:NLead}
\end{equation}
where
\begin{equation}
\mu_{1}=\frac{m_{V}^2}{e^{2}m_{H}^2}
f\left(\frac{m_{V}^{2}}{4\Lambda^{2}}\right).
\label{eqn:NST}
\end{equation}
Here $f(m_{V}^{2}/{4\Lambda^{2}})$ in the next leading string tension
$\mu_{1}$ is a complicated function of $m_{V}^{2}/{4\Lambda^{2}}$ but
finite. As $\Lambda\gg m_{V}^{}$, this function behaves like
$(4\Lambda^2/m_V^2)\{\ln(m_V^2/4\Lambda^2)\}^2$ at most, so that
the ratio of $\mu_{1}$ to $\mu_{0}$ is at most given by
\begin{equation}
\frac{\mu_{1}}{\mu_{0}}\approx\frac{\Lambda^2}{m_H^2}\ln\left(
\frac{\Lambda}{m_V}\right).
\label{eqn:NST/ST}
\end{equation}
It demonstrates that this ratio is small
if $m_{H}^2\gg\Lambda^2\ln(\Lambda/m_V^{})$.
Our large Higgs mass expansion is therefore valid under this condition
on the Higgs mass.
This is consistent with the power counting of (\ref{eqn:PC}),
because $\ln(\Lambda/m_V^{})$ has been counted as $O(1)$ in (\ref{eqn:PC}).

It may be worth while to point out that the calculations for $\mu_0$,
$\alpha_0$ and $\mu_1$ are somewhat similar to the ones for
effective actions and $\beta$ functions in ``the zero-slope limit'' of
(fundamental) string theory or non-linear $\sigma$ models.
More detailed investigation of
this correspondence will be interesting.

To make our discussion clear, let us summarize the conditions used in our
approximation.
In the evaluation of the effective action of the ANO vortex string we
have supposed the following four conditions, under which our
approximation is justified.
(i) The coupling constants are small: $e$, $\lambda<1$.
(This ensures that loop effects are small.)
(ii) The Higgs mass is large enough than the mass of the gauge field:
$m_{H}\gg m_{V}$, meaning $4\la\gg e^{2}$.
(iii) The vortex string is long and smooth enough in comparison with
$1/m_{H}$ and $1/m_{V}$: $m_{H}$, $m_{V}\gg 1/R$.
($R$ denotes the characteristic length which represents the smoothness
of the vortex string.)
(iv) The cut-off parameter $\La$ of the $\de$-function in the
vorticity tensor current is smaller than the Higgs mass and larger than
$1/R$: $m_{H}\gg\La\gg 1/R$.
(The condition $m_{H}\gg\La$ is consistent with the analysis in which
we have used the classical vortex solution as explained under
(\ref{eqn:ANO}).)
Under the conditions (ii), (iii) and (iv), the large Higgs mass
expansion which we have adopted is allowed.
We will also impose the same condition in evaluating the $\FtF$ term
in the next section.
If one can construct a formulation for systematic evaluation of
quantities on the vortex string without imposing the conditions (iii)
and (iv), then one would get more interesting information about the
vortex string.
For example, it would be possible to understand deep relations between
the effective vortex string and the fundamental string.

\section{The $F_{\m\n}\tF_{\m\n}$ term: geometric and topological
properties}\label{sec:FtF}
\reseteqnum

The purpose of this section is to examine geometric and topological
properties of the Abelian Higgs model with vortex strings.
In doing so, our formulation is very useful, since we can treat
the vortex string in arbitrary shape.
To investigate these properties in detail, we concentrate on the
topological term $I$ defined by
\begin{equation}
I=\int\dx \langle\FtF\rangle,
\label{eqn:FtF}
\end{equation}
which in general appears as the chiral anomaly and the CP violating $\th$
term in the action.
Here the definition of the ``dual'' is the following:
$\tF_{\m\n}=\fr{1}{2}\eps F_{\r\s}$.
The expectation value $\langle O \rangle$ indicates the integral value
of $O$ over all fields $V_{\m}$, $\r$, $a_{\m}$ and $B_{\m\n}$ except
the vortex string coordinates $X_{\m}(\s)$ in the path integral
representation.
As will be clarified below, this topological term is closely related to the
geometric and topological properties of the vortex string such as
the self-intersection number of the world sheet swept by the vortex string.
Notice that the world sheet swept by
the vortex string can intersect with itself, that is, the vortex string
coordinate $X_{\m}(\s)$ is an immersion of a two dimensional parameter
space into the four dimensional Euclidean space-time $\bR^{4}$.
In subsequent subsections, first we explicitly evaluate $I$ in the
cases where the number of the vortex strings is conserved and the
vortex string reconnection does not occur, then discuss its generalization.
In addition, in the last part of this section we discuss the chiral
fermion number of the ANO vortex string.

\subsection{Explicit evaluations}\label{sec:f1}

First, we study the case in which a single vortex string exists
and the vortex string reconnection does not occur.
Since the gauge field $A_{\m}$ is given by
$(C_{\m}-a_{\m})/e$ as shown in (\ref{eqn:C}) and
$V_{\m}=C_{\m}-\D_{\m}\omega$, the topological term $I$ can be written as
\begin{equation}
I=\fr{1}{e^{2}}\int\dx
   \langle\ftf-2V_{\m\n}\tf_{\m\n}+V_{\m\n}\tV_{\m\n}\rangle.
\label{eqn:FtF-2}
\end{equation}
We divide $I$ into the following three pieces and evaluate them separately:
\begin{eqnarray}
I^{(1)}\!\!\!&=&\!\!\!\fr{1}{e^{2}}\int\dx\langle\ftf\rangle,\label{eqn:ftf}\\
I^{(2)}\!\!\!&=\!\!\!&-\fr{2}{e^{2}}\int\dx
   \langle V_{\m\n}\tf_{\m\n}\rangle,\label{eqn:Vtf}\\
I^{(3)}\!\!\!&=&\!\!\!\fr{1}{e^{2}}\int\dx\langle V_{\m\n}\tV_{\m\n}\rangle.
\label{eqn:VtV}
\end{eqnarray}

At the beginning, we estimate $I^{(1)}$.
By using $\tf_{\m\n}=-J_{\m\n}/2$, which is derived by variations
of the action (\ref{eqn:TAH}) with respect to $B_{\m\n}$, the
topological term $I^{(1)}$ takes a simpler form
\begin{equation}
I^{(1)}=\fr{1}{4e^{2}}\int\dx J_{\m\n}(x)\tJ_{\m\n}(x).
\label{eqn:JtJ}
\end{equation}
Substituting $J_{\m\n}$ given by (\ref{eqn:Cur}) for this, we obtain
\begin{eqnarray}
I^{(1)}=\fr{4\pi^{2}}{e^{2}}\int\dx\int\dtds\dtdsp
      \Si_{\m\n}(X(\s))\tilde{\Si}_{\m\n}(X(\s'))\nonumber\\
       \qquad\times\fr{\La^{8}}{\pi^{4}}
       \exp\left\{-2\La^{2}x^{2}-\fr{\La^{2}}{2}|X(\s)-X(\s')|^{2}\right\},
\label{eqn:JtJ-2}
\end{eqnarray}
where
\begin{equation}
\Si_{\m\n}(\s)=\D_{1}X_{[\m}(\s)\D_{2}X_{\n]}(\s),
\label{eqn:Area}
\end{equation}
and we have transformed $x_{\m}$ into $x_{\m}+(X_{\m}(\s)+X_{\m}(\s'))/2$.
Integrating over $x$, this becomes
\begin{equation}
I^{(1)}=\fr{4\pi^{2}}{e^{2}}\int\dtds\dtdsp
      \Si_{\m\n}(X(\s))\tilde{\Si}_{\m\n}(X(\s'))
       \Dd(X(\s)-X(\s')),
\label{eqn:AtA}
\end{equation}
where the $\de$-function is regularized as (\ref{eqn:Gauss}).
($\La$ is replaced by $\La/\sqrt{2}$ in the $\de$-function in
(\ref{eqn:AtA}).)
One may worry about the contributions from points $\s_{2}$,
$\s'_{2}=\pm\infty$, since at these points $X_{4}(\s)$ or $X_{4}(\s')$
becomes infinity and then the integral region of $x_{4}$ would not run
from $-\infty$ to $+\infty$.
In order to resolve this difficulty, we introduce a condition
$\D_{2}\bX(\s)=\alpha\D_{1}\bX(\s)$ at $\s_{2}=\pm\infty$, where
$\alpha$ is an arbitrary function of $\s$.
By this condition the integrand of (\ref{eqn:JtJ-2}) becomes zero at
$\s_{2}$, $\s'_{2}=\pm\infty$, so this difficulty disappears.
This condition is not unnatural because it includes adiabatic
processes satisfying $\D_{2}\bX(\s)=0$ at $\s_{2}=\pm\infty$.
We therefore require this condition in this subsection.

Eq. (\ref{eqn:AtA}) shows that this integrand takes non-zero values when
$X(\s)=X(\s')$.
Since we are considering the immersion $X_{\m}(\s)$, there are two
kinds of contributions arising from coincident points where
$X(\s)=X(\s')$ in space-time: the first contribution $I^{(1)}_{1}$
comes from points where $\s=\s'$ in the parameter space and the second
one $I^{(1)}_{2}$ comes from points where $\s\neq\s'$ but $X(\s)=X(\s')$.
$I^{(1)}$ is the sum of $I^{(1)}_{1}$ and $I^{(1)}_{2}$.

Now let us evaluate $I^{(1)}_{1}$ and $I^{(1)}_{2}$ separately.
The procedure of calculation is essentially the same as that of the
previous section.
Using the regularized $\de$-function (\ref{eqn:Gauss}) and the
expansion of $X_{\m}(\s')$ (\ref{eqn:exp-X}), we have the first
contribution
\begin{equation}
I^{(1)}_{1}=-\fr{2\pi^{2}}{e^{2}}PS_{i},
\label{eqn:I11}
\end{equation}
where
\begin{equation}
PS_{i}=\fr{1}{4\pi}\int\dtds\sqrt{g}g^{ab}
    \eps\D_{a}t_{\m\n}\D_{b}t_{\r\s}
\label{eqn:PSi}
\end{equation}
is Polyakov's self-intersection number in which
\begin{equation}
t_{\m\n}=\fr{1}{\sqrt{g}}
     \D_{1}X_{[\m}\D_{2}X_{\n]}.
\label{eqn:tXX}
\end{equation}
Here we have neglected terms which vanish in the $\La\to\infty$ limit.
Indeed, we are considering the case where $1/R\La$ is small, ($R$ is
a characteristic length representing the smoothness of the vortex
string), so that these terms depending on $\La$ are not relevant.
The $PS_{i}$ was discussed from the point of view of the QCD string
involving its extrinsic geometry \cite{Po1,BLS,MN} and is proper to
four dimensional space-time.

Next we will evaluate the second contribution $I^{(1)}_{2}$.
Suppose that the world sheet swept by the vortex string intersects
with itself transversally at many isolated points
$x=p_{1},\dots,p_{n}$: $X(\s_{p_{i}})=X(\s_{p_{i}}')$ at
$\s=\s_{p_{i}}$ and $\s'=\s_{p_{i}}'$, where $\s_{p_{i}}\neq\s_{p_{i}}'$.
In the neighborhood of a ``self-intersection'' point $x=p_{i}$, we can
expand $X_{\m}(\s)$ and $X_{\m}(\s')$ around $\s=\s_{p_{i}}$ and
$\s'=\s_{p_{i}}'$ respectively:
\begin{eqnarray}
X_{\m}(\s)=X_{\m}(\s_{p_{i}})+w_{a}\D_{a}X_{\m}(\s_{p_{i}})+\dots,\nonumber\\
X_{\m}(\s')=X_{\m}(\s_{p_{i}}')+w_{a}'\D_{a}X_{\m}(\s_{p_{i}}')+\dots,
\label{eqn:exp-X-2}
\end{eqnarray}
where $w=\s-\s_{p_{i}}$ and $w'=\s'-\s_{p_{i}}'$.
Using the fact that at the point $x=p_{i}$
\begin{equation}
\Dd(X(\s)-X(\s'))
 =\fr{2}{|\Si_{\m\n}(X(\s_{p_{i}}))\tilde{\Si}_{\m\n}(X(\s'_{p_{i}}))|}
 \de^{(2)}(w)\de^{(2)}(w'),
\label{eqn:Delt}
\end{equation}
we find that a local contribution to $I^{(1)}_{2}$ at $x=p_{i}$ is
$16\pi^{2}SI_{n}(p_{i})/e^{2}$, where
\begin{equation}
SI_{n}(p_{i})
   ={\rm sign}[\Si_{\m\n}(X(\s_{p_{i}}))\tilde{\Si}_{\m\n}(X(\s'_{p_{i}}))]
\label{eqn:SIn}
\end{equation}
is a local intersection number at $x=p_{i}$, namely a suitable sign
determined by the relative orientation at $x=p_{i}$.
Here we have neglected terms which vanish in the $\La\to\infty$ limit.
The total contribution to $I^{(1)}_{2}$ is the sum of local contributions
at all $p_{i}$'s:
\begin{equation}
I^{(1)}_{2}=\fr{16\pi^{2}}{e^{2}}SI_{n},
\label{eqn:I12}
\end{equation}
where
\begin{equation}
SI_{n}=\sum_{i=1}^{n}SI_{n}(p_{i})
\label{eqn:SIn-2}
\end{equation}
is the self-intersection number of the world sheet swept by the vortex
string.

Thus, collecting (\ref{eqn:I11}) and (\ref{eqn:I12}), we get
\begin{equation}
I^{(1)}=-\fr{2\pi^{2}}{e^{2}}PS_{i}+\fr{16\pi^{2}}{e^{2}}SI_{n}.
\label{eqn:I1}
\end{equation}

Next we explain that $I^{(2)}$ and $I^{(3)}$ do not contribute to $I$.
At the leading order of powers of $1/m_{H}$, $I^{(2)}$ and $I^{(3)}$
are respectively given by
\begin{eqnarray}
&&I^{(2)}=\fr{2}{e^{4}}\int\dx d^{4}y\eps
        f_{\m\n}(x)f_{\alpha\beta}(y)\D^{x}_{\r}\D^{x}_{\alpha}
        \langle V_{\s}(x)V_{\beta}(y)\rangle_{0},\label{eqn:I2}\\
&&I^{(3)}=\fr{4}{e^{2}}\int\dx d^{4}y d^{4}z
      \eps f_{\alpha\beta}(y)f_{\ga\de}(z)\D^{y}_{\m}\D^{y}_{\alpha}
        \langle V_{\n}(x)V_{\beta}(y)\rangle_{0}\nonumber\\
&&\hspace{5cm} \times\D^{z}_{\r}\D^{z}_{\ga}
         \langle V_{\s}(x)V_{\de}(z)\rangle_{0}.
\label{eqn:I3}
\end{eqnarray}
We can compute $I^{(2)}$ in the same way to evaluate the effective
action and show that the leading term in powers of $1/\La$, which is
proportional of $PS_{i}$ and $SI_{n}$, becomes zero and remaining parts
vanishes in the $\La\to\infty$ limit .
The difficulty at $\s_{2}$, $\s_{2}'=\pm\infty$ which was mentioned in
evaluating $I^{(1)}$ also appears, but this can be removed by imposing
the same condition.
To show $I^{(3)}=0$ exactly, it is convenient to rewrite the
right-hand side of (\ref{eqn:I3}) in the following:
\begin{equation}
I^{(3)}=4\int\dx d^{4}y d^{4}z\eps
      f_{\alpha\n}(y)f_{\ga\s}(z)
       \D^{y}_{\m}\D^{y}_{\alpha}D(x-y)
       \D^{z}_{\r}\D^{z}_{\ga}D(x-z),
\label{eqn:I3-2}
\end{equation}
where $D(x-y)$ is written as (\ref{eqn:Pr-D}).
Furthermore, using $E(x-y)$ defined as
\begin{equation}
E(x-y)=\int\fr{d^{4}k}{(2\pi)^{4}}\fr{1}{(k^{2}+m_{V}^{2})^{2}}e^{ik(x-y)},
\label{eqn:Ex}
\end{equation}
$I^{(3)}$ becomes
\begin{equation}
I^{(3)}=4\int d^{4}y d^{4}z\eps
       f_{\alpha\n}(y)f_{\ga\s}(z)
       \D^{y}_{\m}\D^{y}_{\alpha}\D^{y}_{\r}\D^{y}_{\ga}
       E(z-y).
\label{eqn:I3-3}
\end{equation}
{}From this we can easily see $I^{(3)}=0$.

After all, $I$ is given by
\begin{equation}
I=-\fr{2\pi^{2}}{e^{2}}PS_{i}+\fr{16\pi^{2}}{e^{2}}SI_{n}.
\label{eqn:I}
\end{equation}
This indicates that the expectation value of the $\FtF$ term becomes
the sum of Polyakov's self-intersection number and the
self-intersection number of the world sheet swept by the vortex string.
The interesting relation between the geometric and topological
quantities $I$, $PS_{i}$ and $SI_{n}$ may be useful in studying the
role of the $\th$ term in systems with the vortex string.
Note that $PS_{i}$ and $SI_{n}$ are different types of
self-intersection numbers.

The next issue is to examine the relation between two kinds of
self-intersection numbers ($PS_{i}$ and $SI_{n}$) and geometric and
topological quantities defined at the boundary of the world sheet where
$\s_{2}=\pm\infty$.
(For simplicity, we have set the initial ``time'' for $-\infty$ and
the final ``time'' for $+\infty$.)
First, as will be proved in appendix A, we can find that $PS_{i}$ is
written as the difference between the initial total twist number and
the final one:
if we take a gauge such that $\s_{2}=X_{4}=t$ and assume that
$\dot{\bX}=\alpha\bX'$ at $t=\pm\infty$, then we get
\begin{equation}
PS_{i}=-4T_{w}(\bX(t);\bn(t))\bigg|^{t=+\infty}_{t=-\infty},
\label{eqn:PSi-Tw}
\end{equation}
where
\begin{equation}
T_{w}(\bX(t);\bn(t))=
    \fr{1}{2\pi}\oint d\s_{1}[\be (\s_{1},t)\times \bn(\s_{1},t)]
       \cdot \bn'(\s_{1},t)
\label{eqn:Tw}
\end{equation}
is the total twist number in which $\be=\bX'/|\bX'|$ and the prime
denotes the differentiation by $\s_{1}$.
Here $\bn(\s_{1},t)$ is a smoothly varying unit vector which is
perpendicular to $\bX'$ at each point and is a periodic function of
$\s_{1}$ with period $2\pi$:
$\bn^{2}=1$, $\bn\cdot\bX'=0$ and $\bn(\s_{1}+2\pi,t)=\bn(\s_{1},t)$.
This total twist number is a geometric quantity dependent on $\bX$
locally and on the topological class of $\bn$, but not a topological
invariant.
If we modify $\bn$ globally, then the total twist
number changes by some integer.
This ambiguity will be settled when we consider the self-linking number
together, as will be discussed later.
Note that this total twist number can be interpreted as the spin
factor in three dimensions, which is related to the torsion of a path
\cite{Po2}.

We next consider the relation between the self-intersection number and the
self-linking number.
The self-linking number $SL_{k}(\bX(t);\bn(t))$ at $t$ is defined as
the linking number of a vortex string $\bX(\s_{1},t)$ and a fictitious
vortex string $\bX(\s_{1},t)+\ve\bn(\s_{1},t)$, where $\bn$ is equal to
the one used in the definition of the total twist number (see
fig. \ref{fig:SLK}).
This self-linking number is independent of $\ve$ and is a
topological invariant which can take only integer values.
\epsfhako{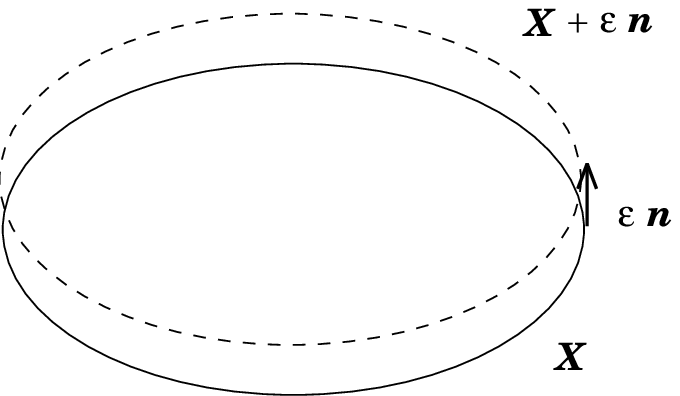}{t}{7cm}{The solid line and the dotted line
denote a vortex string $\vec{X}$ and a fictitious one $\vec{X}+\ve\vec{n}$,
respectively. The self-linking number of $\vec{X}$ is defined
as the linking number of them. In this figure, the self-linking number
is zero.}{fig:SLK}{}
One can think of this ``framing'' as a thickening of the vortex string
into a ribbon bounded by $\bX$ and $\bX+\ve\bn$.
As shown in appendix B, the intersection number of two world sheets
swept by two different vortex strings coincides with the difference
between the linking number at $t=-\infty$ and that at $t=+\infty$.
When the vortex string $\bX$ crosses the fictitious one $\bX+\ve\bn$ at
a crossing point in a projecting plane, as illustrated in
fig. \ref{fig:SIS}, they crosses twice each other at the point.
\epsfhako{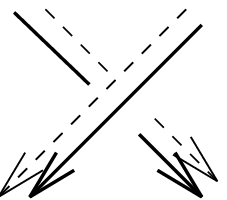}{t}{2.5cm}{The vortex string (the solid
line) crosses the fictitious one (the dotted line) twice at a crossing
point in a projecting plane.}{fig:SIS}{}
Thus, the self-linking number turns out to be twice the linking
number of $\bX$ and $\bX+\ve\bn$.
On the other hand, the self-intersection number of the vortex string
$X_{\m}(\s)$ is equal to the intersection number of $X_{\m}(\s)$ and
$X_{\m}+\ve n_{\m}(\s)$ as $\ve\to 0$.
Therefore, we obtain
\begin{equation}
SI_{n}=-\fr{1}{2}SL_{k}(\bX(t);\bn(t))\biggl|^{t=+\infty}_{t=-\infty}
     \qquad:
\label{eqn:SIn-SLk}
\end{equation}
the self-intersection number of $X_{\m}(\s)$ is the half of the
difference of the self-linking number at $t=-\infty$ and the one at
$t=+\infty$.

Using the relations between the quantities defined in the
``bulk'' and those defined at the ``boundary'' as in
 (\ref{eqn:PSi-Tw}) and (\ref{eqn:SIn-SLk}), we obtain
\begin{equation}
I=-\fr{8\pi^{2}}{e^{2}}
 \Bigl[SL_{k}(\bX(t);\bn(t))
  -T_{w}(\bX(t);\bn(t))\Bigr]^{t=+\infty}_{t=-\infty}.
\label{eqn:SLk-Tw}
\end{equation}
Furthermore, as was shown in \cite{Ca} (see also \cite{FKV}), the
combination
$SL_{k}-T_{w}$ is free from the ambiguity of the definition of $\bn$
and is called the writhing number $W_{r}$:
\begin{equation}
SL_{k}(\bX(t);\bn(t))-T_{w}(\bX(t);\bn(t))=W_{r}(\bX(t)),
\end{equation}
where
\begin{equation}
W_{r}(\bX(t))=\fr{1}{4\pi}\e_{ijk}\oint_{\Ga}dx_{i}\oint_{\Ga}dy_{i}
              \fr{(\bx-\by)_{k}}{|\bx-\by|^{3}}.
\label{eqn:defWr}
\end{equation}
Here $\Ga$ denotes the configuration of the vortex string
$\bX(\s_{1},t)$ (see also appendix B).
Consequently, we can represent $I$ by using the writhing number:
\begin{equation}
I=-\fr{8\pi^{2}}{e^{2}}[W_{r}(\bX(+\infty))-W_{r}(\bX(-\infty))].
\label{eqn:DWr}
\end{equation}
When the vortex string intersects with itself transversally, the
writhing number changes by 2 before and after the time of intersection.
Note that the writhing number is
a geometric quantity, which is a continuous function except at the
time of intersection.

Although we have considered the case with a single vortex string as
yet, it is easy to extend the result on a single vortex string to the
case with many vortex strings.
In the case with $n$ vortex strings $X_{p}(\s)$ ($p=1,\cdots,n$), there
appears a new contribution to $I$ given by
\begin{equation}
\fr{4\pi^{2}}{e^{2}}\sum_{p,q=1 \atop (p\neq q)}^{n}\int\dtds\dtdsp
               \Si_{\m\n}(X_{p}(\s))\tilde{\Si}_{\m\n}(X_{q}(\s'))
               \Dd(X_{p}(\s)-X_{q}(\s')).
\end{equation}
By using the same argument in evaluating the self-intersection number,
this contribution can be computed to be the sum of intersection
numbers between different vortex strings:
\begin{equation}
\fr{8\pi^{2}}{e^{2}}\sum_{p,q=1 \atop (p\neq q)}^{n}I_{n}(X_{p},X_{q}),
\end{equation}
where $I_{n}(X_{p},X_{q})$ denotes the intersection number of two
different vortex strings $X_{p}$ and $X_{q}$.
Furthermore, as shown in appendix B, the intersection number is
represented as the difference between the linking number at
$t=-\infty$ and that at $t=+\infty$, so we get
\begin{equation}
I=-\fr{8\pi^{2}}{e^{2}}\left[\sum_{p=1}^{n}W_{r}(\bX_{p}(t))
 +\sum_{p,q=1 \atop (p\neq q)}^{n}L_{k}(\bX_{p}(t),\bX_{q}(t))\right]
 ^{t=+\infty}_{t=-\infty}.
\label{eqn:Wr+Lk}
\end{equation}
The first term comes from ``self-interactions'' of vortex strings and
the second one comes from ``mutual interactions'' between different
vortex strings.
Note that the writhing number serves as a measure of the right-left
asymmetry of the vortex string, that is a measure of its chirality
\cite{FKV}.

\subsection{Generalization}\label{sec:f2}

Until now, we have considered the case in which the creation,
annihilation and reconnection of the vortex string do not occur.
However, if we adopt several natural assumptions, we can get more general
results which include our results (\ref{eqn:Wr+Lk}) and the
ones in the cases where the number of vortex strings
is not conserved.
The first key to the generalization is the observation that in
the previous subsection the massive gauge field $C_{\m}$ (or $V_{\m}$)
had no effect in evaluating $I$ at least at the leading order.
It is therefore perhaps natural to assume that this is the case even in more
general cases at the leading order.
The second key is that the $\FtF$ term can be rewritten as a total
divergence: $\FtF=\eps\D_{\m}(A_{\n}F_{\r\s})$.
We assume as usual that $\eps A_{\n}F_{\r\s}$ is continuous and
$F_{\m\n}$ becomes zero in the spatial infinity, so that we get
\begin{equation}
I=-\int d^{3}x \langle\epsi A_{i}F_{jk}\rangle
     \bigg |^{t=+\infty}_{t=-\infty}.
\label{eqn:CS}
\end{equation}
Furthermore, using the first assumption that $C_{\m}$ (or $V_{\m}$)
does not contribute to $I$ and the fact that
$A_{\m}=(C_{\m}-a_{\m})/e$, $I$ can be written as
\begin{equation}
I=-\fr{1}{e^{2}}\int d^{3}x\epsi a_{i}f_{jk}\bigg|^{t=+\infty}_{t=-\infty}.
\label{eqn:CSV}
\end{equation}
As an example, let us consider the case where no vortex string exists
at the initial time and only one vortex string $\bX$ exists at the
final time.
Since $a_{i}$ satisfies $f_{\m\n}=-2\tilde{J}_{\m\n}$, $a_{i}$ is
easily solved in the Coulomb gauge $\D_{i}a_{i}=0$ at the final time
$t=+\infty$:
\begin{equation}
a_{i}(\bx,t=+\infty)=\fr{1}{2}\epsi\oint_{\Ga}dy_{j}
                    \fr{(\bx-\by)_{k}}{|\bx-\by|^{3}},
\label{eqn:a}
\end{equation}
where $\Ga$ denotes the position of the vortex string
$\bX(\s_{1},t=+\infty)$.
Here we have taken the $\La\to\infty$ limit .
As a result, we obtain
\begin{equation}
I=-\fr{8\pi^{2}}{e^{2}}W_{r}(\bX(t=+\infty)),
\label{eqn:I-Wr}
\end{equation}
where the path of the integral in this writhing number ranges over the
position of the vortex string at the final time $\bX(\s_{1},t=+\infty)$.
This result (\ref{eqn:I-Wr}) can be easily extended to more general
cases where $m$ vortex strings exist at the initial time and $n$
vortex strings exist at the final time:
\begin{equation}
I=-\fr{8\pi^{2}}{e^{2}}\Bigl[W_{r}(t)+L_{k}(t)\Bigr]^{t=+\infty}_{t=-\infty}
\label{eqn:I-Wr+Lk}
\end{equation}
where $W_{r}(t)$ and $L_{k}(t)$ denote the sum of the writhing numbers
of each vortex string at $t$ and the sum of the linking numbers
between vortex strings at $t$, respectively (see (\ref{eqn:Blur})).
Taking $n=m$, one can easily see that this generalized result
(\ref{eqn:I-Wr+Lk}) includes the previous result (\ref{eqn:Wr+Lk})
which has evaluated exactly at least at the leading order. We can
examine the chiral fermion number of the ANO vortex string by using
 (\ref{eqn:I-Wr+Lk}) or (\ref{eqn:DWr}), as will be shown in the
next subsection.

It is worth emphasizing that the explicit evaluation of $I$ such as
(\ref{eqn:I}), (\ref{eqn:DWr}) and (\ref{eqn:Wr+Lk}) is important and
necessary itself because through it we can find several interesting
relations between geometric or topological quantities and without
ambiguities we can realize the cases in which the creation,
annihilation and reconnection of the vortex string do not occur.
The explicit evaluation is of course necessary to justify the generalization
where several assumptions are made.

\subsection{The chiral fermion number of a closed ANO vortex
string}\label{sec:f3}
It was pointed out that if topological
defects are coupled to fermions, they might have fermion numbers \cite{JR}.
In this subsection, using the results obtained in sect. \ref{sec:f1} and
sect. \ref{sec:f2}, we derive the chiral fermion number of the ANO
vortex string in arbitrary shape.

We consider models with anomalous global $U(1)$ symmetries and local
$U(1)$ symmetries which are spontaneously broken.
Although our consideration is model-independent, for definiteness,
let us consider the following model defined by
\begin{equation}
{\cal L}=\frac{1}{4}\FF
    +|(\D_{\m}-ieA_{\m})\varphi|^{2}
    +2\la\biggl(\varphi^{\dagger}\varphi+\fr{\eta^{2}}{2}\biggr)^{2}
    +i\bar{\psi}(\sD-ie\sA)\psi.
\label{eqn:AH+D}
\end{equation}
This Lagrangian has a chiral symmetry $\psi\to e^{i\gamma_{5}\th}\psi$
which is anomalous, so that the conservation law of the
chiral current becomes
\begin{equation}
\D_{\m}\langle j^{5}_{\m}\rangle
 =-i\fr{e^{2}}{8\pi^{2}}\langle F_{\m\n}\tF_{\m\n}\rangle,
\label{eqn:Cons}
\end{equation}
where $j^{5}_{\m}=\bar{\psi}\ga_{\m}\ga_{5}\psi$.
Thus the chiral fermion number
\begin{equation}
Q_{5}=-i\int d^{3}x \langle j^{5}_{4}\rangle
\label{eqn:Q5}
\end{equation}
satisfies
\begin{equation}
Q_{5}(+\infty)-Q_{5}(-\infty)
=-\int\dx [\nabla_{i}\langle j^{5}_{i}\rangle
  +\fr{e^{2}}{8\pi^{2}}\langle F_{\m\n}\tF_{\m\n}\rangle].
\label{eqn:DQ5}
\end{equation}
In order to calculate the chiral fermion number of a closed ANO
vortex string, we consider a field configuration which starts at the
trivial vacuum when $t=-\infty$ and arrives at a closed ANO vortex
string configuration when $t=+\infty$.
We assume that the field configuration at spatially
infinity is topologically trivial all the time.
This is possible because we are considering a closed vortex string.
As a result the first term of the right-hand side of (\ref{eqn:DQ5}) does not
contribute to $Q_{5}$.
Substituting (\ref{eqn:I-Wr+Lk}) for the second
term of the right-hand side of (\ref{eqn:DQ5}), we get
\begin{equation}
Q_{5}(+\infty)=W_{r}(\bX),
\label{eqn:Q5-2}
\end{equation}
where $\bX$ denotes the position of the vortex string at $t=+\infty$.
In general, $Q_{5}(+\infty)$ has two kinds of contributions: the first
comes from the fermion production and the second comes from the chiral
fermion number of the ANO vortex string.
However, since the first contribution
affects only integer part of $Q_{5}$ and $W_{r}(\bX)$ can take
non-integer values, we can conclude that the closed
ANO vortex has a non-zero chiral fermion number related to the
writhing number:
\begin{equation}
Q_{5}^{{\rm vortex}}=W_{r}(\bX)\qquad({\rm mod}\>\bZ).
\label{eqn:Q5V}
\end{equation}

Here we make some comments. (i) To derive (\ref{eqn:Q5V}), we have used
the relation (\ref{eqn:I-Wr+Lk}) which needs several assumptions.
If we only use the explicitly evaluated result (\ref{eqn:DWr}), we obtain
$\Delta Q_{5}^{{\rm vortex}}=\Delta W_{r}(\bX)\>\>({\rm mod}\>\bZ)$.
Namely, the change of the chiral fermion number of the ANO vortex
string is equal to the change of its writhing number. Thus
$Q_{5}^{{\rm vortex}}=W_{r}(\bX)+const\>\>({\rm mod}\>\bZ)$,
where ``$const$'' does not depend on the shapes of the vortex string.
(ii) We have imposed the condition ``$\D_{2}\bX=\alpha\D_{1}\bX$ at
$t=\pm\infty$'' to obtain (\ref{eqn:DWr}), but in evaluating the
chiral fermion number this condition is automatically satisfied since
in this case we have only to treat the vortex string which moves
adiabatically \cite{GW}.

There may be interesting phenomena when the vortex string passes
through itself, because the writhing number $W_{r}$ suffers
discontinuities and changes by $\pm 2$ at that time.
Anyway, we realize that the shapes of the vortex string must be important to
investigate the anomalous fermion production through the ANO vortex string.

\section{The vortex string in a model with a broken global $U(1)$
symmetry}{\label{sec:Glo}}
\reseteqnum

The application of our formulation is not restricted to only the Abelian
Higgs model.
Using the ``topological'' formulation, we can also treat the vortex
string in a model with a broken global $U(1)$ symmetry.
Although this model is simpler than the Abelian Higgs model,
some differences appear, which are briefly sketched in this section.
The Lagrangian with a broken global $U(1)$ symmetry in the Euclidean
formulation is given by
\begin{equation}
{\cal L}=\D_{\m} \varphi ^{\dagger} \D_{\m} \varphi
  +2\la\biggl(\varphi^{\dagger}\varphi-\fr{\eta^{2}}{2}\biggr)^{2},
\label{eqn:Go}
\end{equation}
where $\varphi$ is a complex scalar field.
As explained in sect. 2, it is convenient to use the following
Lagrangian in considering the dynamics of the vortex string:
\begin{equation}
{\cal L}=|(\D_{\m}-ia_{\m})\phi|^{2}
  +2\la\biggl(\phi^{\dagger}\phi-\fr{\eta^{2}}{2}\biggr)^{2}
   +i\Bf+iB_{\m\n}J_{\m\n}.
\label{eqn:TGo}
\end{equation}
Here $a_{\m}$ is a $U(1)$ gauge field for the vorticity, $B_{\m\n}$ a rank
two antisymmetric tensor field, $f_{\m\n}$ a field strength tensor of
$a_{\m}$.
The definitions of the complex scalar field $\phi$ and the vorticity tensor
current $J_{\m\n}$ are given by (\ref{eqn:Tra}) and
 (\ref{eqn:Cur}), respectively.
This Lagrangian is equivalent to (\ref{eqn:Go}) in the same sense that
 (\ref{eqn:TAH}) is
equivalent to (\ref{eqn:AH}).

In order to estimate the effective action of the vortex string, we
take the unitary gauge and replace the fields as follows:
\begin{eqnarray}
&&\phi(x)=\fr{1}{\sqrt{2}}
     (\eta+\r(x))e^{i\omega(x)},\label{eqn:rho-2}\\
&&U_{\m}(x)=a_{\m}(x)-\D_{\m}\omega(x).
\label{eqn:U}
\end{eqnarray}
Thus the Lagrangian becomes
\begin{eqnarray}
{\cal L}&=&\fr{1}{2}\D_{\m}\r\D_{\m}\r+2\la\eta^{2}\r^{2}
           +\fr{1}{2}\eta^{2}U_{\m}U_{\m}\nonumber\\
        & &+\eta\r U_{\m}U_{\m}+\fr{1}{2}\r^{2}U_{\m}U_{\m}
         +2\la\eta\r^{3}+\fr{1}{2}\la\r^{4}\nonumber\\
        & &+i\ep B_{\m\n}U_{\r\s}+iB_{\m\n}J_{\m\n},
\label{eqn:UGo}
\end{eqnarray}
where $U_{\m\n}=\D_{\m}U_{\n}-\D_{\n}U_{\m}$.
At first sight the propagating mode looks to be only the $\r$ field,
but the field $U_{\m}$ contains a massless
mode implicitly, as will be explained below.
The evaluation of the effective action can be performed as in the
Abelian Higgs model.
As before the propagator of the $\r$ field is given by
\begin{equation}
\langle\r (x)\r(y)\rangle_{0}=\int\fr{d^{4}k}{(2\pi)^{4}}
              \fr{1}{k^{2}+m_{H}^{2}}e^{ik(x-y)},
\label{eqn:Pr-r-2}
\end{equation}
where $m_{H}^{2}=4\la\eta^{2}$.
Furthermore, we can use the large mass expansion by
the powers of $1/m_{H}$.
After integrating over the $\r$ field, we get at the tree level
\begin{equation}
{\cal L}=\fr{1}{2}{\eta^{2}}U_{\m}U_{\m}+i\ep B_{\m\n}U_{\r\s}
        +iB_{\m\n}J_{\m\n}+\cdots,
\label{eqn:G-Lead}
\end{equation}
where ``$\cdots$'' in (\ref{eqn:G-Lead}) denote terms with more than
one piece of $U_{\m}U_{\m}$.
Using ordinary relations between numbers of vertices, propagators and
external lines for diagrams, one can find that the terms with $N$ pieces
of $U_{\m}U_{\m}$ are suppressed by $2(N-1)$-th powers of $1/m_{H}$.
Therefore, if $U_{\m}$ satisfies a condition $U_{\m}\ll m_{H}$, we can
perform a systematic approximation by using the large Higgs mass
expansion.
In the last part of this section, we will show that this condition is
satisfied if cut-off parameters in the present model are smaller
enough than $m_{H}$.

Now let us explain how a massless mode, which corresponds to the
Goldstone mode, is derived from the above Lagrangian (\ref{eqn:G-Lead}).
Integrating over $B_{\m\n}$, the partition function $Z$ becomes the
following form:
\begin{equation}
Z=\int{\cal D}U_{\m}\de(\ep U_{\r\s}+J_{\m\n})
\exp(-\int d^{4}x\fr{1}{2}\eta^{2}U_{\m}U_{\m}+\cdots),
\label{eqn:DU}
\end{equation}
where ``$\cdots$'' in (\ref{eqn:DU}) are the same terms as in
(\ref{eqn:G-Lead}).
To integrate over $U_{\m}$, the constraint
$\ep U_{\r\s}+J_{\m\n}=0$ must be solved to be
\begin{eqnarray}
&&U_{i}(x)=\fr{1}{8\pi}\int
         d^{3}y\fr{1}{|\bx-\by|}\epsi\D_{j}^{y}J_{4k}(y)
         +\fr{\sqrt{2}}{\eta}\D_{i}\pi (x),\label{eqn:Ui}\\
&&U_{4}(x)=\fr{1}{16\pi}\int
         d^{3}y\fr{1}{|\bx-\by|}\epsi\D_{i}^{y}J_{jk}(y)
         +\fr{\sqrt{2}}{\eta}\D_{4}\pi (x),\label{eqn:U0}
\end{eqnarray}
where $\pi(x)$ is an arbitrary function and $y_{4}=x_{4}$.
This arbitrary function $\pi(x)$ reflects the ``gauge symmetry'' of the above
constraint for $U_{\m}$.
However $\pi(x)$ cannot be fixed through this ``gauge symmetry''
because the integrand $\exp(-\int
d^{4}x\eta^{2}U_{\m}U_{\m}/2+\cdots)$ in (\ref{eqn:DU}) is not gauge
invariant.
Indeed, after integrating over $U_{\m}$, we get
\begin{equation}
Z=\int{\cal D}\pi\exp(-\int d^{4}x \D_{\m}\pi\D_{\m}\pi+\cdots).
\label{eqn:Dpi}
\end{equation}
One can easily find that the terms ``$\cdots$'' in (\ref{eqn:Dpi})
do not contain a quadratic term of $\pi(x)$, so $\pi(x)$ represents a
massless physical mode.

At the leading order of the large Higgs mass expansion, the Lagrangian
(\ref{eqn:UGo}) is reduced to a more familiar form, which can be written as
\begin{equation}
{\cal L}_{0}=\fr{1}{2}{\eta^{2}}U_{\m}U_{\m}
    +i\ep B_{\m\n}U_{\r\s}+i\BJ.
\label{eqn:bf-leading}
\end{equation}
If we first integrate over $U_{\m}$ in (\ref{eqn:bf-leading}), then the
Lagrangian becomes
\begin{equation}
{\cal L}_{0}=\frac{4}{3\eta^{2}}H_{\m\n\la}H_{\m\n\la}
                 +iB_{\m\n}J^{\m\n},
\label{eqn:KR}
\end{equation}
where
\begin{equation}
H_{\m\n\la}=\D_{\m}B_{\n\la}+\D_{\la}B_{\m\n}
                  +\D_{\n}B_{\la\m}.
\end{equation}
Thus in the large Higgs mass limit, this model gives rise to the
Kalb-Ramond model coupling to the vortex string through $J_{\m\n}$ \cite{KR}.
This fact was pointed out by other authors \cite{DS}.
In the total action obtained in \cite{DS}, however, there appears a
term with the inverse of $|\phi|$, which is proportional to
$1/|\phi|^{2}\cdot H_{\m\n\r}H_{\m\n\r}$, so that an infinit number of
vertices of $\r$ comes out in the perturbative expansion around the
non-zero expectation value $\eta$ of the Higgs field.
On the other hand, in our formulation where the starting Lagrangian is
(\ref{eqn:UGo}) (or (\ref{eqn:TGo})), there appears only a finite number of
vertices of $\r$, so that it is easy to perform the perturbative
expansion in a systematic way.
This advantage of our formulation becomes clearer in the evaluation of
higher order corrections.
In addition, it is also remarkable that we have two different
descriptions of the model (\ref{eqn:Dpi}) and (\ref{eqn:KR}), depending
on the order of integration over $U_{\m}$ and $B_{\m\n}$ in
(\ref{eqn:G-Lead}) or (\ref{eqn:UGo}).
This can be regarded as a kind of ``dual transformations''.

Finally, we comment on the cut-off parameters which are necessary for
justifying the large Higgs mass expansion.
In order to satisfy $U_{\m}\ll m_{H}$, we have to introduce two
types of cut-off parameters:
the first is the cut-off $\La$ for the $\de$-function in $J_{\m\n}$ which
was introduced in the ANO vortex string case (see
 (\ref{eqn:Gauss})) and the second is the ``cut-off'' restricting
the momentum of the massless mode $\pi(x)$.
Now let us imagine a circle $C$ which is on a plane perpendicular to a
tangent direction at a point on the vortex string and has a radius
$r$ from that point.
When $r>1/\La$, the first terms in (\ref{eqn:Ui}) and
(\ref{eqn:U0}) roughly coincide with $-\D_{\m}\theta$ ($\theta$ is a
solid angle subtended by the vortex string) and this $\D_{\m}\theta$
takes almost the same value on the circle $C$, thus satisfying
$\oint_{C}dx^{\m}\D_{\m}\theta=2\pi$ and
$\oint_{C}dx^{\m}\D_{\m}\theta\approx\D_{\m}\theta\times 2\pi r$.
As a result, we find $\D_{\m}\theta\approx 1/r$ on the circle $C$.
On the other hand, when $r<1/\La$, the first terms in
(\ref{eqn:Ui}) and (\ref{eqn:U0}) are smaller than $\La$ on the
circle $C$ because we are regularizing the $\de$-function in $J_{\m\n}$
by the cut-off $\La$: for example, in
the case of a straight vortex string, the first term at $r=0$ becomes
zero.
Putting both cases together, the first terms turn out to be smaller
than $\La$ in the whole region.
Furthermore, since the momentum of $\pi(x)$ is limited by the second
``cut-off'', the second terms in (\ref{eqn:Ui}) and (\ref{eqn:U0}) are
smaller than the second ``cut-off''.
Therefore, if the cut-off parameters are smaller enough than $m_{H}$,
the condition $U_{\m}\ll m_{H}$ is satisfied.

\section{Conclusions}

In the preceding sections, we have developed the ``topological''
formulation which allows the systematic analysis of the effective
vortex string in arbitrary shape and have applied to the Abelian Higgs
model and the model with a broken global $U(1)$ symmetry.
Using our formulation, in particular, we have evaluated the effective
action of the vortex string and the expectation value of the
topological $\FtF$ term.
As a result, many geometric and topological quantities concerning the
vortex string have been derived.
{}From the effective action of the ANO vortex string including the
Nambu-Goto term and the extrinsic curvature squared term with negative
sign, one can realize the motion of the vortex string, which indicates
that the vortex string prefers curving as far as it is smooth enough.
Furthermore, we have found that the ANO vortex string has a non-zero chiral
fermion number related to the writhing number (modulo $\bZ$) and
have suggested that interesting phenomena such as the anomalous
fermion production might occur through intersection processes of the ANO
vortex string.
It should be emphasized that the chiral fermion number of the ANO
vortex string depends on its ``shape''.
In addition, we have shown remarkable relations between $I$, $PS_{i}$,
$SI_{n}$, $T_{w}$, $SL_{k}$ and $W_{r}$ in (\ref{eqn:I}), (\ref{eqn:PSi-Tw}),
(\ref{eqn:SIn-SLk}), (\ref{eqn:SLk-Tw}),
(\ref{eqn:DWr}), (\ref{eqn:Wr+Lk}) and (\ref{eqn:I-Wr+Lk}).
They must be useful themselves in studying geometric or topological
properties of the vortex string and the role of the $\th$ term.

In our ``topological'' formulation, there have appeared non-zero
extrinsic quantities of the ANO vortex string such as the extrinsic
curvature squared term, Polyakov's self-intersection number and the
writhing number.
On the other hand, it was argued that there appears no extrinsic
curvature squared term in the effective action of the ANO vortex
string in the formulation based on F\"orster's parameterization of
coordinates \cite{Gr}, which we call the F\"orster-Gregory (FG)
formulation.
Furthermore, if one evaluates the topological $\FtF$ term by using
the FG formulation, this term turns out to be
zero at least at the leading order.
This is due to the fact that the electric field vanishes in the
static ANO vortex solution, which is applied at the leading order in
the FG formulation.
Note that $\FtF\approx\bB\cdot\bE$, where $\bB$ is a magnetic field
and $\bE$ an electric field.
In our formulation, as we have said, we have found the
$\FtF$ term to take a non-zero value at the leading order.
It is not easy to compared the FG formulation with
ours, because in the FG formulation the equation of motion is used
instead of the path integral representation which we have adopted.
However, these discrepancies between the FG formulation and ours likely come
from the differences between the parameterizations of coordinates and
the regularizations of the vortex core in each formulation.

We would like to stress the following points.
(i) In our formulation, the Lorentz invariance and the conservation of
the vorticity are manifestly satisfied all the time.
Indeed, we have used the Lorentz invariant Guassian-type
regularization for the $\de$-function in the vorticity tensor current.
Furthermore, the static ANO vortex solution, which is not Lorentz
invariant at first sight, is not used at all.
(ii) Our perturbative calculation (e.g. by the large Higgs mass
expansion) is systematic and efficient.
In addition, the path integral representation is convenient to
evaluate physical quantities on the vortex string in a systematic
manner.
(iii) Our formulation can be applied to the model with a broken global
$U(1)$ symmetry as shown in sect. 4, while it is difficult to adopt the
FG formulation for that model.
(iv) Using our formulation, the dynamics of quantized vortices in
superfluid can be examined \cite{HYAT}.
In this case, the effective action of the vortex string turns out to
be of the same form as the action of a vortex string in an
incompressible perfect fluid, so that we can explain the phenomena in
experiments on the quantized vortex by applying our formulation.
On the grounds mentioned above, our ``topological'' formulation is
reliable.

Our formulation can be directly applied to superconductor
systems, the cosmic string model and grand unified models with extra
$U(1)$ symmetries.
In particular, it is of interest to investigate the possibility of the
fermion number violation by the vortex string in more detail in
various cases including the Weinberg-Salam theory.
It will be also suggestive to examine strings and gravity from the
point of view of the effective string.

\vskip1cm
\centerline{\large\bf Acknowledgements}
We would like to thank our colleagues in
Kyoto University for encouragement.

\appendix
\newpage
\noindent
\begin{center}
{\Large {\bf Appendices}}
\end{center}

\section{Relation between Polyakov's self-intersection number and
the total twist number}
\reseteqnum

In this appendix A, we derive the relation (\ref{eq:a0}) under the
assumption $\dot{\bX}=\alpha\bX'$ at $t=\pm\infty$,
where $\alpha$ is some arbitrary function of $(\s_{1},t)$:
\begin{equation}
\fr{1}{16\pi}\int\dtds\sqrt{g}g^{ab}
    \eps\D_{a}t_{\m\n}\D_{b}t_{\r\s}
=-T_{w}(\bX(t);\bn(t))\bigg|^{t=+\infty}_{t=-\infty}.
\label{eq:a0}
\end{equation}
Here $t_{\m\n}=\D_{1}X_{[\m}\D_{2}X_{\n]}/\sqrt{g}$ and
the total twist number $T_{w}$ is defined as
\begin{equation}
T_{w}(\bX(t);\bn(t))=
\fr{1}{2\pi}\oint d\s_{1}[\be(\s_{1},t)\times\bn(\s_{1},t)]
   \cdot\bn'(\s_{1},t),
\end{equation}
where $\be=\bX'/|\bX'|$.
Since the assumption $\dot{\bX}=\alpha\bX'$ at $t=\pm\infty$ coincides
with the condition $\D_{2}\bX=\alpha\D_{1}\bX$ at $t=\pm\infty$ which
has been required in sect. 3.1 (see under (\ref{eqn:AtA})), it is not
unnatural.

Let us first rewrite the left-hand side of (\ref{eq:a0}) by the
extrinsic curvature.
The extrinsic curvature of the world sheet $X^{\m}(\s)$ is
defined as
\begin{equation}
\D_{a}\D_{b}X_{\m}=\Ga^{c}_{ab}\D_{c}X_{\m}+K^{A}_{ab}n_{\m}^{A}.
\label{eq:a1}
\end{equation}
Here $\Ga^{c}_{ab}$ is the Christoffel symbol in terms of derivatives
of $g_{ab}$ and $n^{A}_{\m}$ is a unite vector satisfying
$n^{A}_{\m}n^{B}_{\m}=\de^{AB}$ and $\D_{a}X_{\m}n_{\m}^{A}=0$.
In addition, we choose such a direction of $n^{A}_{\m}$ as $\eps
n^{1}_{\m}n^{2}_{\n}\D_{1}X_{\r}\D_{2}X_{\s}>0$.
Using (\ref{eq:a1}), $\eps\D_{a}X_{\m}\D_{b}X_{\n}\D_{c}X_{\r}=0$ and $\eps
n^{A}_{\m}n^{B}_{\n}\D_{a}X_{\r}\D_{b}X_{\s}=\sqrt{g}\>\e^{AB}\e_{ab}$,
we obtain
\begin{equation}
\fr{1}{16\pi}\int\dtds\sqrt{g}g^{ab}
    \eps\D_{a}t_{\m\n}\D_{b}t_{\r\s}
=-\fr{1}{4\pi}\int\dtds g^{ab}\e^{AB}\e_{cd}K^{A}_{ac}K^{B}_{bd}.
\label{eq:a2}
\end{equation}
There is another expression of the right-hand side of (\ref{eq:a2}):
\begin{equation}
-\fr{1}{4\pi}\int\dtds g^{ab}\e^{AB}\e_{cd}K^{A}_{ac}K^{B}_{bd}
=-\fr{1}{4\pi}\int\dtds\e^{AB}\e_{ab}\D_{a}(n_{\m}^{A}\D_{b}n_{\m}^{B}).
\label{eq:a3}
\end{equation}
The above relation is derived by using
$\e^{AB}\e_{ab}(n^{A}_{\m}\D_{a}n^{C}_{\m})(n^{B}_{\n}\D_{b}n^{C}_{\n})=0$
and
\begin{equation}
\D_{a}n^{A}_{\m}
=-(n^{A}_{\s}\D_{a}n^{B}_{\s})n^{B}_{\m}-K^{A}_{ab}g^{bc}\D_{c}X_{\m}.
\label{eq:a6}
\end{equation}
Eq. (\ref{eq:a6}) is proved easily by (\ref{eq:a1}) and the
completeness of the vectors
$\D_{a}X_{\m}$ and $n^{A}_{\m}$.
Combining (\ref{eq:a2}) and (\ref{eq:a3}), we get \cite{MN}
\begin{equation}
\fr{1}{16\pi}\int\dtds\sqrt{g}g^{ab}
    \eps\D_{a}t_{\m\n}\D_{b}t_{\r\s}
=-\fr{1}{4\pi}\int\dtds\e^{AB}\e_{ab}\D_{a}(n_{\m}^{A}\D_{b}n_{\m}^{B}).
\label{eq:a4}
\end{equation}
We can construct $n_{\m}^{A}$ explicitly by using a vector
$n_{\m}=(\bn,0)$ satisfying $\bn^{2}=1$, $\bX'\cdot\bn=0$ and
$\bn(\s_{1}+2\pi,t)=\bn(\s_{1},t)$:
\begin{eqnarray}
n^{1}_{\m}&=&\fr{1}{\sqrt{1-\fr{(n\cdot\D_{2}X)^{2}g_{11}}{g}}}
             \left[n_{\m}-\fr{(n\cdot\D_{2}X)}{g}
              (g_{11}\D_{2}X_{\m}-g_{12}\D_{1}X_{\m})\right],\nonumber\\
n^{2}_{\m}&=&-\fr{1}{2}\eps n^{1}_{\n}t_{\r\s}.
\label{eq:a5}
\end{eqnarray}
If $g\neq0$ and $\bn^{2}\neq0$, $g-(n\cdot\D_{2}X)^{2}g_{11}$ is not
zero, so the above $n^{A}_{\m}$ are well-defined.
One can easily see that these $n^{A}_{\m}$ satisfy the conditions such
that $n^{A}_{\m}n^{B}_{\m}=\de^{AB}$, $\D_{a}X_{\m}n_{\m}^{A}=0$ and
$\eps n^{1}_{\m}n^{2}_{\n}\D_{1}X_{\r}\D_{2}X_{\s}>0$.
Since $\bn$ and $X_{\m}$ are periodic functions of $\s_{1}$, $n^{A}_{\m}$
also become periodic functions of $\s_{1}$.
Thus the integral of
$\e^{AB}\D_{1}(n^{A}_{\m}\D_{2}n^{B}_{\m})$ vanishes, and the
right-hand side of (\ref{eq:a4}) becomes
\begin{equation}
\fr{1}{2\pi}\oint d\s_{1}(n^{1}_{\m}\D_{1}n^{2}_{\m})
 \bigg|^{\ta=+\infty}_{\ta=-\infty}.
\label{eqn:SDN}
\end{equation}
Substituting (\ref{eq:a5}) for (\ref{eqn:SDN}), the integrand
$n^{1}_{\m}\D_{1}n^{2}_{\m}$
becomes
\begin{equation}
\fr{1}{\sqrt{g}\>(1-\fr{(n\cdot\D_{2}X)^{2}g_{11}}{g})}\eps n_{\m}
 \left[\D_{1}n_{\n}-\fr{n\cdot\D_{2}X}{g}
  (g_{11}\D_{1}\D_{2}X_{\n}-g_{12}\D_{1}\D_{1}X_{\n})\right]
  \D_{1}X_{\r}\D_{2}X_{\s}.
\end{equation}
Now we take $\s_{2}=X_{4}=t$ gauge.
By the assumption $\dot{\bX}=\alpha\bX'$ at $t=\pm\infty$,
$n\cdot\D_{2}X$ vanishes  and $g$ becomes $|\bX'|^{2}$ as $t\to\pm\infty$.
Therefore, we obtain
\begin{equation}
\fr{1}{2\pi}\oint d\s(n^{1}_{\m}\D_{1}n^{2}_{\m})
 \bigg|^{t=+\infty}_{t=-\infty}
=\fr{1}{2\pi}\oint d\s_{1}[\be(\s_{1},t)\times\bn(\s_{1},t)]
   \cdot\bn'(\s_{1},t)\bigg|^{t=+\infty}_{t=-\infty}.
\label{eqn:a10}
\end{equation}
{}From (\ref{eqn:a10}) and (\ref{eq:a4}), (\ref{eq:a0}) holds.

\section{Relation between the intersection number and the linking number}
\reseteqnum

In this appendix B, we explain the relation between the intersection
number and the linking number.

There are many methods for defining the linking number, all of which
are equivalent.
Our definition in this paper is the following (see \cite{Ro}).
Let $J$ and $K$ be two disjoint oriented knots in $\bR^{3}$, which are
corresponding to vortex strings at a fixed time $t$.
Consider a ``regular'' projection of $J$ and $K$ to any plane.
At each point $c_{i}$ ($i=1,\cdots,m$) where $J$ crosses $K$ in the
projecting plane, ${\rm sign}(c_{i})$ is defined as $+1$ for the case
depicted in the left picture of fig. \ref{fig:LK} and $-1$ for the
case depicted in the right one of fig. \ref{fig:LK}.
\epsfhako{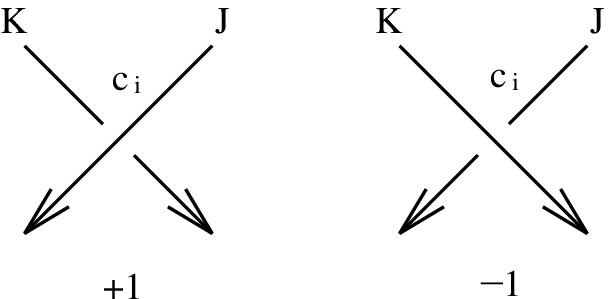}{htb}{6cm}{The definition of sign($c_{i}$). $J$
crosses $K$ at a point $c_{i}$ in a projecting plane.}{fig:LK}{}
The linking number of $J$ and $K$, $L_{k}(J,K)$, is defined by the half
of the sum of all signs at $c_{i}$'s:
\begin{equation}
L_{k}(J,K)=\fr{1}{2}\sum_{i=1}^{m}{\rm sign}(c_{i}).
\label{eqn:B-Lk}
\end{equation}
The linking number satisfies $L_{k}(J,K)=L_{k}(K,J)$ and
$L_{k}(-J,K)=-L_{k}(J,K)$ where $-J$ is $J$ with the inverse
orientation.

Next consider two different world sheets $X_{\m}(\s)$ and
$Y_{\m}(\s')$ which intersect transversally at $p_{i}$
($i=1,\cdots,n$).
We take a gauge such that $\s_{2}=X_{4}$ and $\s'_{2}=Y_{4}$, and
identify $X_{4}$ and $Y_{4}$ as ``time'': $X_{4}=Y_{4}=t$.
Let us look at the neighborhood of a intersection point
$p=(\bp,p_{4})$ among $p_{i}$'s.
(We can get the total result by collecting the local contributions at
$p_{i}$'s.)
If two vortex strings move from a state such as the left picture of
fig. \ref{fig:IS} to a state such as the right one of fig.
\ref{fig:IS} as time goes from $t<p_{4}$ to $t>p_{4}$, the linking
number of $\bX(\s_{1},t)$ and $\bY(\s'_{1},t)$ changes by $-1$:
\begin{equation}
L_{k}(\bX,\bY;t>p_{4})-L_{k}(\bX,\bY;t<p_{4})=-1,
\end{equation}
where $L_{k}(\bX,\bY;t)$ indicates the linking number of $\bX$ and
$\bY$ at $t$.
\epsfhako{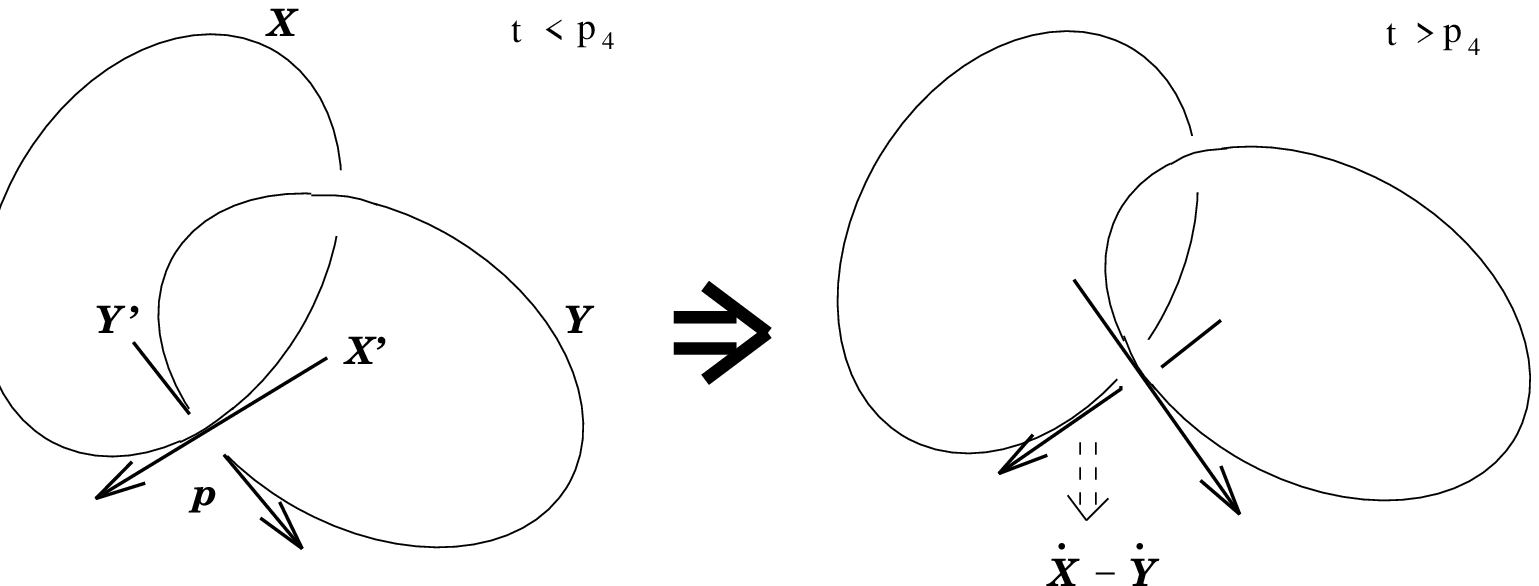}{htb}{15.5cm}{A example of the two different
world sheets $X_{\m}(\s)$ and $Y_{\m}(\s')$ which intersect transversally
at $(\vec{p},p_{4})$. The configurations of $\vec{X}(\s_{1},t)$ and
$\vec{Y}(\s_{1}',t)$ at $t<p_{4}$ are illustrated in the left picture,
and the ones at $t>p_{4}$ in the right picture. The solid arrows
denote the tangent vectors $\vec{X}'$ and $\vec{Y}'$, and the dotted
arrow denotes the vector $\dot{\vec{X}}-\dot{\vec{Y}}$.}{fig:IS}{}

On the other hand, in our gauge, the local intersection number of
$X_{\m}(\s)$ and $Y_{\m}(\s')$ at $p$
is given by
\begin{equation}
I_{n}(p)={\rm sign}[\Si_{\m\n}(X)\tilde{\Si}_{\m\n}(Y)]
={\rm sign}[\epsi X_{i}'Y_{j}'(\dot{Y_{k}}-\dot{X_{k}})].
\end{equation}
When the location of two vortex strings changes from the left picture
to the right one in fig. \ref{fig:IS}, then the above sign becomes $+1$,
namely the intersection number at $p$ is $+1$.
Therefore, it turns out that the local intersection number at $p$
coincides with the difference between the linking numbers of $\bX$ and
$\bY$ before and after the time $t=p_{4}$:
\begin{equation}
I_{n}(p)=-L_{k}(\bX,\bY;t)\biggl|^{t>p_{4}}_{t<p_{4}}.
\end{equation}
Repeating the consideration above at each local intersection point
$p_{i}$, we obtain
\begin{equation}
I_{n}[X,Y]=-L_{k}(\bX,\bY;t)\biggl|^{t=+\infty}_{t=-\infty},
\end{equation}
where
\begin{equation}
I_{n}[X,Y]=\sum_{i=1}^{n}I_{n}(p_{i})
\end{equation}
is the intersection number of $X_{\m}(\s)$ and $Y_{\m}(\s')$.
We conclude that the intersection number of the world sheets is equal
to the difference between the linking number at $t=-\infty$ and that
at $t=+\infty$.

On the other hand, the linking number of $J$ and $K$ defined by
(\ref{eqn:B-Lk}) is equivalent to the Gauss linking number, which can
be written as
\begin{equation}
\fr{1}{4\pi}\epsi\oint_{J}dx_{i}\oint_{K}dy_{j}
             \fr{(\bx-\by)_{k}}{|\bx-\by|^{3}},
\label{eqn:Blur}
\end{equation}
where $\bx$ ranges over $J$ and $\by$ over $K$.
Eq. (\ref{eqn:Blur}) has used in sect. 3.2.
Assuming $J=K$, this Gauss linking number becomes the writhing number
of $J$.

Finally, we comment on the writhing number $W_{r}$ defined by
(\ref{eqn:defWr}).
Let us parameterize $\bx$ and $\by$ ranging over the path $\Ga$ in
(\ref{eqn:defWr}) as $\bX(\s_{1})$ and $\bX(\s_{1}')$ respectively, where
$0\le\s_{1}\le 2\pi$ and $0\le\s_{1}'\le 2\pi$.
Furthermore, define the domain
$D=\{\s'_{1}|\s_{1}-\de\le\s'_{1}\le\s_{1}+\de\}$ where $\de$ is a small
constant, and then expand $\bX(\s_{1}')$ around $\s_{1}'=\s_{1}$.
Inserting the expanded $\bX(\s_{1}')$ into the numerator and the
denominator in the integrand of the writhing number (\ref{eqn:defWr}),
we find that the numerator and the denominator behave like
$|\s_{1}'-\s_{1}|^{4}$ and $|\s_{1}'-\s_{1}|^{3}$ respectively.
The behavior $|\s_{1}'-\s_{1}|^{4}$ of the numerator is due to the
antisymmetric property of the $\e$ tensor.
Therefore, the writhing number is not singular in the region where
$\bX(\s_{1})=\bX(\s_{1}')$ (i.e. $\bx=\by$) since the integrand of the
writhing number behaves like $|\s_{1}'-\s_{1}|$ in $D$, that is in the
domain where $\s_{1}'\approx\s_{1}$.
Here we have assumed that the path $\Ga$ does not intersect with itself.

\newcommand{\J}[4]{{\sl #1} {\bf #2} (19#3) #4}
\newcommand{\MPL}{Mod.~Phys.~Lett.}
\newcommand{\NP}{Nucl.~Phys.}
\newcommand{\PL}{Phys.~Lett.}
\newcommand{\PR}{Phys.~Rev.}
\newcommand{\PRL}{Phys.~Rev.~Lett.}
\newcommand{\AP}{Ann.~Phys.}
\newcommand{\CMP}{Commun.~Math.~Phys.}
\newcommand{\CQG}{Class.~Quant.~Grav.}
\newcommand{\PRP}{Phys.~Rept.}
\newcommand{\SPU}{Sov.~Phys.~Usp.}
\newcommand{\RMPA}{Rev.~Math.~Pur.~et~Appl.}
\newcommand{\SPJ}{Sov.~Phys.~JETP}


\begin{thebibliography}{99}
\bibitem{NO} H.B.~Nielsen and P.~Olesen, \J{\NP}{B61}{73}{45}.
\bibitem{Fo} D.~F\"{o}rster, \J{\NP}{B81}{74}{84}.
\bibitem{GS} J.L.~Gervais and B.~Sakita, \J{\NP}{B91}{75}{301}.
\bibitem{HYAT} M.~Hatsuda, S.~Yahikozawa, P.~Ao and D.J.~Thouless,
              \J{\PR}{B49}{94}{15870}.
\bibitem{BBRT} D.~Birmingham, M.~Blau, M.~Rakowski and G.~Thompson
             \J{\PRP}{209}{91}{129}.
\bibitem{Na} Y.~Nambu, \J{\NP}{B130}{77}{505}.
\bibitem{Va} T.~Vachaspati, \J{\PRL}{68}{92}{1977}.
\bibitem{KO} F.R.~Klinkhamer and P.~Olesen, ``A new perspective on
             electroweak strings'', preprint NIKHEF-H/94-02, hep-ph/9402207.
\bibitem{tH} G.~'t~Hooft, \J{\PRL}{37}{76}{8}.
\bibitem{KRS} V.A.~Kuzmin, V.A.~Rubakov and M.E.~Shaposhinikov,
              \J{\PL}{155B}{85}{36}.
\bibitem{Vi} A.~Vilenkin, \J{\PRP}{121}{85}{263}.
\bibitem{WZ} F.~Wilczek and A.~Zee, \J{\PRL}{51}{83}{2250}.
\bibitem{LNS}W.~Lerche, B.E.W.~Nilsson and A.N.~Schellekens,
            \J{\NP}{B289}{87}{609}.
\bibitem{CDJM}R.~Capovilla, J.~Drell, T.~Jacobson and L.~Manson,
             \J{\CQG}{8}{91}{41}.
\bibitem{Po1} A.~Polyakov, \J{\NP}{B268}{86}{406}.
\bibitem{Ab} A.A.~Abrikosov, \J{\SPJ}{5}{57}{1174}.
\bibitem{BLS} A.P.~Balachandran, F.~Lizzi and G.~Sparano,
             \J{\NP}{B263}{86}{608}.
\bibitem{MN} P.O.~Mazur and V.P.~Nair, \J{\NP}{B284}{86}{146}.
\bibitem{Po2} A.M.~Polyakov, \J{\MPL}{A3}{88}{325}.
\bibitem{Ca} G.~Calagareanu, \J{\RMPA}{4}{59}{5};
             \J{Czwch.~Math.~J.}{11}{61}{588}.
\bibitem{FKV} M.D.~Frank-Kamenetski\u{\i} and A.V.~Vologodski\u{\i},
              \J{\SPU}{24}{81}{679}.
\bibitem{JR} R.~Jackiw and C.~Rebbi, \J{\PR}{D13}{76}{3398}.
\bibitem{GW} J.~Goldstone and F.~Wilczek, \J{\PRL}{47}{81}{986}.
\bibitem{KR} M.~Kalb and P.~Ramond, \J{\PR}{D9}{74}{2273}.
\bibitem{DS} R.L.~Davis and E.P.S.~Shellard, \J{\PL}{214B}{88}{219}.
\bibitem{Ro} D.~Rolfsen, Knots and Links (Publish or Perish, 1976).
\bibitem{Gr} R.~Gregory, \J{\PR}{D43}{91}{520}.




\end{thebibliography}
\end{document}